\documentclass{aa}

\usepackage[flushleft]{threeparttable}
\usepackage{graphicx}
\usepackage{xtab,afterpage}
\usepackage{amsmath}
\usepackage{epstopdf}
\usepackage{url}
\usepackage[flushleft]{threeparttable}
\usepackage{booktabs,fixltx2e}
\usepackage{amssymb}
\usepackage{textcomp}
\usepackage{hyperref} 
\usepackage{multirow} 
\usepackage{times}
\usepackage{txfonts}

\renewcommand{\v}{\ensuremath{\mathbf{v}}}

\makeatletter
\newcommand*{\rom}[1]{\expandafter\@slowromancap\romannumeral #1@}
\makeatother

\title{Stellar winds can affect gas dynamics in debris disks and create observable belt winds}\titlerunning{Presence of belt winds in extrasolar systems}
\author{Quentin Kral\inst{1}\thanks{E-mail: quentin.kral@obspm.fr} \and J. E. Pringle\inst{2} \and Luca Matr{\`a}\inst{3} \and Philippe Thébault\inst{1}}

\institute{LESIA, Observatoire de Paris, Universit{\'e} PSL, CNRS, Sorbonne Universit{\'e}, Univ. Paris Diderot, Sorbonne Paris Cit{\'e}, 5 place Jules Janssen, 92195 Meudon, France\\
\and
Institute of Astronomy, University of Cambridge, Madingley Road, Cambridge CB3 0HA, UK\\
\and
School of Physics, Trinity College Dublin, The University of Dublin, College Green, Dublin 2, Ireland\\
}

\begin{document}

   \date{Received September 15, 1932; accepted March 16, 1937}



\label{firstpage}


  \abstract
   {Gas is now detected in many extrasolar systems around mature stars aged between 10 Myr to $\sim$ 1 Gyr with planetesimal belts. Gas in these mature disks is thought to be released from planetesimals and has been modelled using a viscous disk approach where the gas expands inwards and outwards from the belt where it is produced. Therefore, the gas has so far been assumed to be a circumstellar disk orbiting the star but at low densities, this may not be a good assumption as the gas could be blown out by the stellar wind instead.}
   {In this paper, we aim to explore when the transition from a gas disk to such a gas wind happens and whether it can be used to determine the stellar wind properties around main-sequence stars that are otherwise hard to measure.}
   {We developed an analytical model for A to M stars that can follow the evolution of gas outflows and target when the transition occurs between a disk or a wind to finally compare to current observations. The crucial criterion is here the gas density for which gas particles stop being protected from stellar wind protons impacting at high velocities on radial trajectories.}
   {We find that: 1) Belts of radial width $\Delta R$ with gas densities $< 7 \, (\Delta R/50 {\rm \, au})^{-1}$ cm$^{-3}$ would create a wind rather than a disk, which would explain the recent outflowing gas detection in NO Lup. 2) The properties of this belt wind can be used to measure stellar wind properties such as their densities and velocities. 3) Very early-type stars can also form gas winds because of the star's radiation pressure rather than stellar wind. 4) Debris disks with low fractional luminosities $f$ are more likely to create gas winds, which could be observed with current facilities.}
   {The systems containing low gas masses such as Fomalhaut or TWA 7 or more generally, debris disks with fractional luminosities $f \lesssim 10^{-5} (L_\star/L_\odot)^{-0.37} $ or stellar luminosity $\gtrsim 20 \, L_\odot$ (A0V or earlier) would rather create gas outflows (or belt winds) than gas disks. Gas observed to be outflowing at high velocity in the young system NO Lup could be an example of such belt winds. Future observing predictions in this wind region should account for the stellar wind to be able to detect the gas. The detection of these gas winds is possible with ALMA (CO and CO$^+$ could be good wind tracers) and would allow us to constrain the stellar wind properties of main-sequence stars, which are otherwise difficult to measure (e.g. there are no successful measures around A stars for now).}


   \maketitle
   
\section{Introduction}
Gas is now detected in most dense planetesimal belts (observed as bright debris disks) around young early-type stars $>10$ Myr \citep{2017ApJ...849..123M}. It is now also detected around up to Gyr-old stars \citep{2017ApJ...842....9M,2017MNRAS.465.2595M} and around later-type stars, all the way from A-to-M stars \citep[e.g.,][]{2016MNRAS.460.2933M,2019AJ....157..117M,2020MNRAS.497.2811K,2022MNRAS.509..693R}, with a remarkable diversity of CO gas masses ranging from 0.1 \citep[e.g.][]{2013ApJ...776...77K,2017ApJ...849..123M,2019ApJ...884..108M} to $10^{-7}$ M$_\oplus$ \citep{2017ApJ...842....9M}. The observed CO gas and its daughter products (C and O) are best described as being secondary \citep{2017MNRAS.469..521K,2019MNRAS.489.3670K}, i.e., the gas is released from planetesimals. It is only for the few most massive systems that a primordial origin (i.e. the hypothesis that the gas would be a remnant of the protoplanetary disk phase) is not completely ruled out. However, there are strong indications that, even for these massive systems, the observed gas is of secondary origin \citep{2017ApJ...839...86H,2021arXiv211107655S} and CO remains abundant thanks to shielding by carbon naturally produced in a secondary fashion as explained in detail in \citet{2019MNRAS.489.3670K}.
  
These discoveries have prompted several numerical investigations aimed at understanding the origin and evolution of this long-lived gas component. Up to now, this gas has been modelled as a circumstellar disk orbiting the star and mostly co-located with the planetesimal belts. In these models, the gas production rate has been assumed to be proportional to the dust mass loss rate of the planetesimal belt. This modelling approach was applied to $\sim 200$ systems and it can explain most observations to date \citep{2017MNRAS.469..521K}. Two exceptions to the standard scenario are given by the detection of an atomic gas wind in $\eta$ Tel in UV \citep{2021AJ....162..235Y} and probably in optical \citep{2018A&A...614A...3R} and also around $\sigma$ Her in UV \citep{2003ApJ...582..443C}. The central stars being very early ($\sim$A0V), ionised carbon may also become unbound and cannot retain other atomic species through braking Coulomb collisions as is usually assumed \citep{2006ApJ...643..509F}, which is similar to the results of \citet[][Fig.~11]{2017MNRAS.469..521K}.

\begin{figure*}
   \centering
   \includegraphics[width=15.5cm]{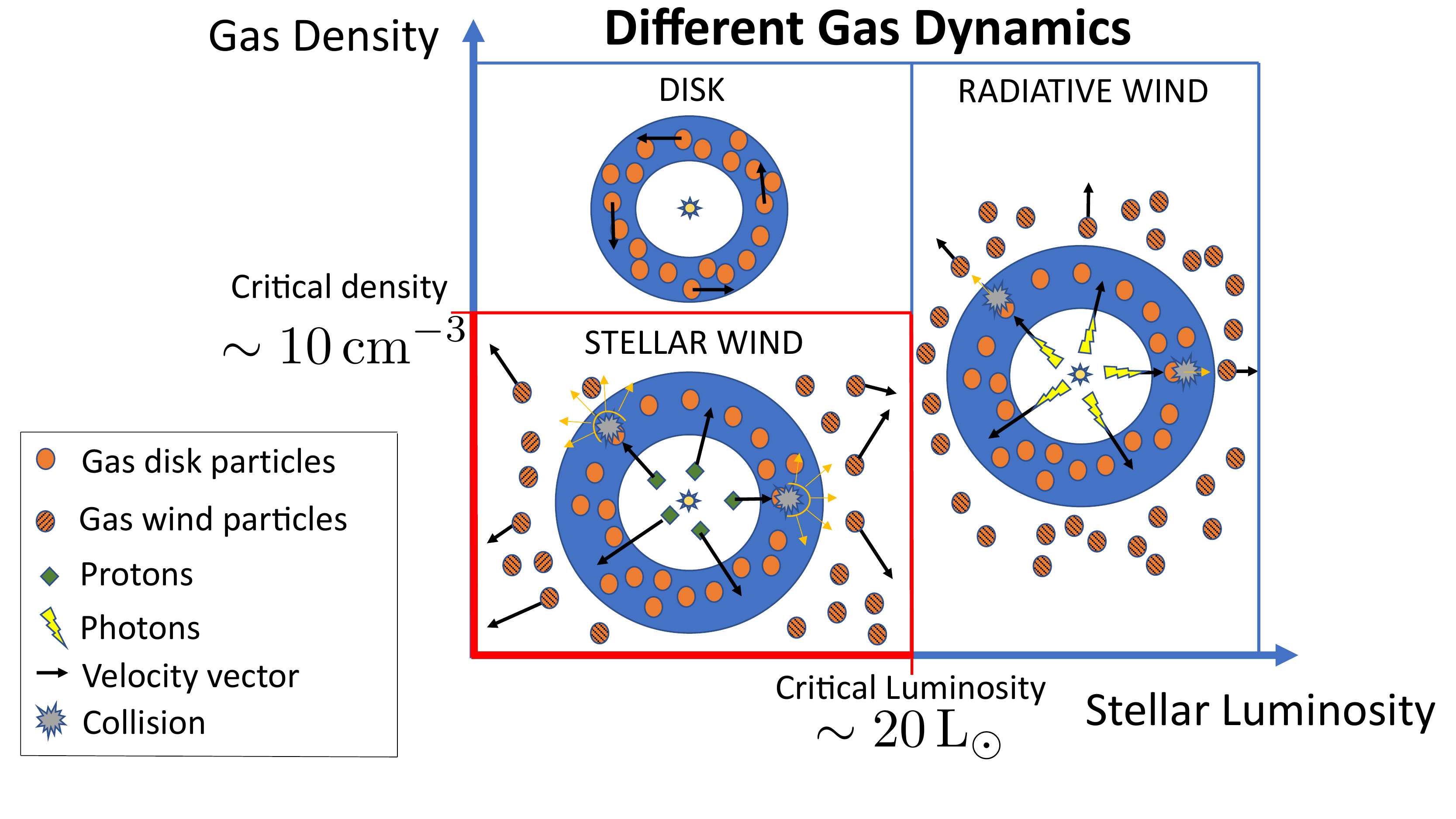}
   \caption{\label{figcartoon}  Cartoon showing the different dynamics of the gas as a function of gas density and stellar luminosity. The critical density and luminosity showed on the figure are approximate and more precise values are presented in the paper. This paper focuses on the Stellar wind mechanism circled in red.}
\end{figure*}

In addition, when recently studying the low gas environment of the Solar System Kuiper Belt (KB), \citet{2021A&A...653L..11K} found that the possible secondary gas released (because of progressive internal warming of large planetesimals) by KBOs (Kuiper belt objects) could be directly blown out by the Solar wind without forming a disk-like structure.
Indeed, the likely low gas density in the KB would mean that gas particles are not protected from the Solar wind and each wind proton hitting a gas particle will lead to an ejection (at higher than the local escape velocity) of the gas particle and hence create a gas belt wind. This KBO gas would have a lower density than expected by current viscous gas models and a high velocity heading outwards, characteristic of a wind. Here, the mechanism driving the wind would be stellar wind (SW) and not stellar radiation as presented in the previous paragraph for $\eta$ Tel. SWs can only affect low gas density systems while radiative winds can impact any gas (that remains optically thin to high energy photons) given that the star is roughly earlier than A0V, which does not apply to many debris disks. The cartoon presented in Fig.~\ref{figcartoon} shows the classical picture of a gas disk in Keplerian rotation along with the stellar wind and radiative wind mechanisms that become important at low gas densities and high stellar luminosities, respectively.

In this paper, we aim to trace the behaviour of low density gas in presence of SWs and generalize the current models to any circumstellar gaseous system. Our main goal is, in particular, to explore what are the crucial criteria that will determine if the behaviour of a given gas system will be disk-like or wind-like and whether those ``belt'' winds can be detected by current instruments. To do so, we developed an analytical model to describe the gas density and velocity in low gas mass systems where the effect of SWs can become important. We will also investigate whether these belt winds can be used as proxies to determine the SW properties around main-sequence stars that are otherwise hard to measure \citep{2015A&A...577A..27J}, especially for A stars where no measurements led to a detection so far \citep{1992A&A...257..663L,2014A&A...564A..70K}. We will also explore the type of debris disks in terms of fractional luminosity and stellar type that could harbour these belt winds to be able to target them with, e.g., ALMA.
 
\section{The analytical model}

We will describe the gas as an idealised one zone model with a scale height H and a constant density throughout for a planetesimal belt located between $R$ and $R+\Delta R$. 

The stellar wind velocity around M to A stars is in the range 100-1000 km/s \citep[corresponding to the range of escape velocities at the stellar surface,][]{2015A&A...577A..27J}. After an elastic collision with a wind proton of velocity $v_{\rm SW}$ at an angle $\psi$ (between the proton velocity vector and the normal to the surfaces of proton and gas particle spheres at the point of contact), a gas particle of mean molecular weight $\mu$ will have a velocity equals to 

\begin{equation}\label{vgas}
v_g=\frac{2 \cos(\psi) v_{\rm SW}}{\mu +1}
\end{equation}

\noindent provided that the initial gas particle velocity (commonly of a few km/s along the azimuth) is small compared to the end velocity after impact (see Appendix \ref{coloutcome} to get full expressions). Assuming that the wind velocity is close to the escape velocity at the stellar surface \citep{2015A&A...577A..27J} then we obtain that a gas particle will become unbound (i.e. with a velocity after impact greater than the local escape velocity) for $R>(\mu+1)/(2 \sqrt{2} \cos(\psi)) R_\star$, which we assume in our model. Indeed, for a CO molecule suffering a head-on collision this criterion translates into $R \gtrsim 10 R_\star \sim 0.05$ au, which will be always verified for debris disks (even a worst case scenario of a collision with an impact angle $\psi=89.5$ deg would imply $R \gtrsim 5$ au, which will also be true as belts are typically located at tens of au).

The model we present here is developed for systems with low gas densities where the mean free path of a wind proton $\lambda_{\rm w}$ crossing the belt is much greater than the belt's width $\Delta R$. This means that a wind proton will at most interact with one gas particle in the disk (see cartoon in Fig.~\ref{figcartoon} for the basic mechanism). When $\lambda_{\rm w}$ becomes lower than $\Delta R$, much less gas particles can be ejected and the density starts building up to quickly reach the usual steady-state gas disk regime that has been described in the literature up to now (see Appendix \ref{gasmodel}). The main criterion to check whether a system is in the wind regime is thus

\begin{equation}\label{Kneq}
\lambda_{\rm w} \gg \Delta R.
\end{equation}

If we rewrite this criterion in terms of the Knudsen number $K_{\rm n}=\lambda_{\rm w}/H$, we obtain that $K_{\rm n} \gg \Delta R/H $, and thus that $K_{\rm n} \gg 1$ because the scale height is much smaller than the belt's width even for the narrowest disks. For such high values of $K_{\rm n}$, gas will not behave like a fluid, and one should use a collisional approach to model the dynamics of the gas rather than standard hydrodynamics as it then behaves as the sum of all individual particles.

The mean free path of a wind proton crossing the gas in the belt can be defined as a function of the gas density $n_{\rm g}$ and its elastic cross-section $\sigma_{\rm col}$ as

\begin{equation}\label{mfpeq}
\lambda_{\rm w}=\frac{1}{n_{\rm g} \sigma_{\rm col}},
\end{equation}

\noindent where $\sigma_{\rm col}=\pi R^2_{\rm col}$ depends on the particles that are considered to collide with each other. We find that $R_{\rm col}$ depends on the species considered. For an ionised species such as CO$^+$ or C and O neutral atoms, it is roughly equal to the radius of the species considered, i.e. $\sim 0.78$ \AA \, \citep{1978AdAMP..13....1M,1997CP....223...59O} or a cross-section of $\sim 2 \times 10^{-20}$ m$^2$ (see Appendix \ref{beltmodel}). Considering CO, the elastic cross-section with a high-velocity proton is $\sim 2 \times 10^{-18}$ m$^2$ \citep{1992CP553,2006JChPh.124c4314D}, which leads to $R_{\rm col} \sim 8$ \AA. Solving for Eq.~\ref{Kneq} and using Eq.~\ref{mfpeq} to find the critical gas density $n_{\rm crit}$ below which gas is in the wind regime leads to a gas density

\begin{equation}\label{ncriteq}
n_{\rm g}<n_{\rm crit}=\frac{1}{\Delta R \, \sigma_{\rm col}},
\end{equation}

\noindent which can be turned into



\begin{equation}\label{ncritneutraleq}
n_{\rm crit} \sim 7 \, {\rm cm}^{-3}  \left( \frac{\Delta R}{50 \, {\rm au}} \right)^{-1} \left( \frac{\alpha_X}{\alpha_{\rm CO}} \right)^{-2/3}
\end{equation}

\noindent where $\alpha_X$ is the polarisability of species $X$ (see Appendix \ref{beltmodel}) and we assumed that $R_{\rm col}$ is in the limit where its radius is fixed by the particle's radius (case where collisions happen with ions or C, or O). In the case of protons colliding with CO, $n_{\rm crit}$ is 100 times smaller. In general, we find that proton collisions mostly happen faster than ionisation of the gas released in the belt (see Appendix \ref{ionisation}). Indeed, for a wide range of host stars, interstellar radiation, and also accounting for SW proton ionisation, we find that the ionisation timescale is most often at least 10 times greater than the time for a gas particle to get hit by a proton at the density levels we consider (see Fig.~\ref{fig2}). Therefore, it is most likely that collisions will happen in majority between stellar protons and neutral atoms (e.g. C, O) or molecules (e.g. CO), but this should be checked on a case-by-case basis. We note that it also means that a CO gas disk could be above the critical density but after ionisation of the molecules, CO$^+$ behaves as a wind because of its much smaller cross-section, hence why the most constraining value of the critical density is not necessarily that for CO collisions. Ionisation of O and CO can also happen through charge exchange with SW protons with a cross-section of $\sigma_{\rm exc} \sim 1.3 \times 10^{-19}$ m$^{2}$ \citep{1997A&A...317..193I,2017PhPro..90..391L}. Comparing the charge exchange ionisation timescale (which dominates over stellar ionisation) to that of the collisional timescale for CO, we find that the order of magnitude for the ionisation fraction of CO is of order 0.1.


Now, we define a stellar wind proton mass loss rate $\rm{\dot{M}_{\rm SW}}$ and a gas production rate in the disk $\rm{\dot{M}_g}$. In the case that $\lambda_w \gg \Delta R$, each proton hits $\Delta R/\lambda_w \ll 1$ gas particles as it passes through the disk. The number of protons of mass $m_p$ entering the disk per unit time is then given by ${\rm{\dot{N}_p}} = ({\rm{\dot{M}_{\rm SW}}} / m_p) (H/R)$. Therefore the number of gas particles hit per unit time is ${\rm{\dot{N}_g}}={\rm{\dot{N}_p}}  (\Delta R/\lambda_w)$. If we assume that we are at steady state and that, in addition, each gas particle exits the disk instantaneously once it is hit by a proton, then $\rm{\dot{N}_g}$ must equal the rate at which gas particles are being replenished, i.e. we get ${\rm{\dot{M}_g}}/(\mu m_p)=({\rm{\dot{M}_{\rm SW}}}/m_p) (H/R) (\Delta R/\lambda_w)$. From this we find the gas density at steady state in the wind regime, which is equal to

\begin{equation}\label{ngasss3}
n_{\rm d}=\left(\frac{\rm{\dot{M}_g}}{{\rm{\dot{M}_{\rm SW}}}}\right) \left(\frac{R}{\mu H \Delta R \sigma_{\rm col}}\right).
\end{equation}

\begin{figure}
   \centering
   \includegraphics[width=9.5cm]{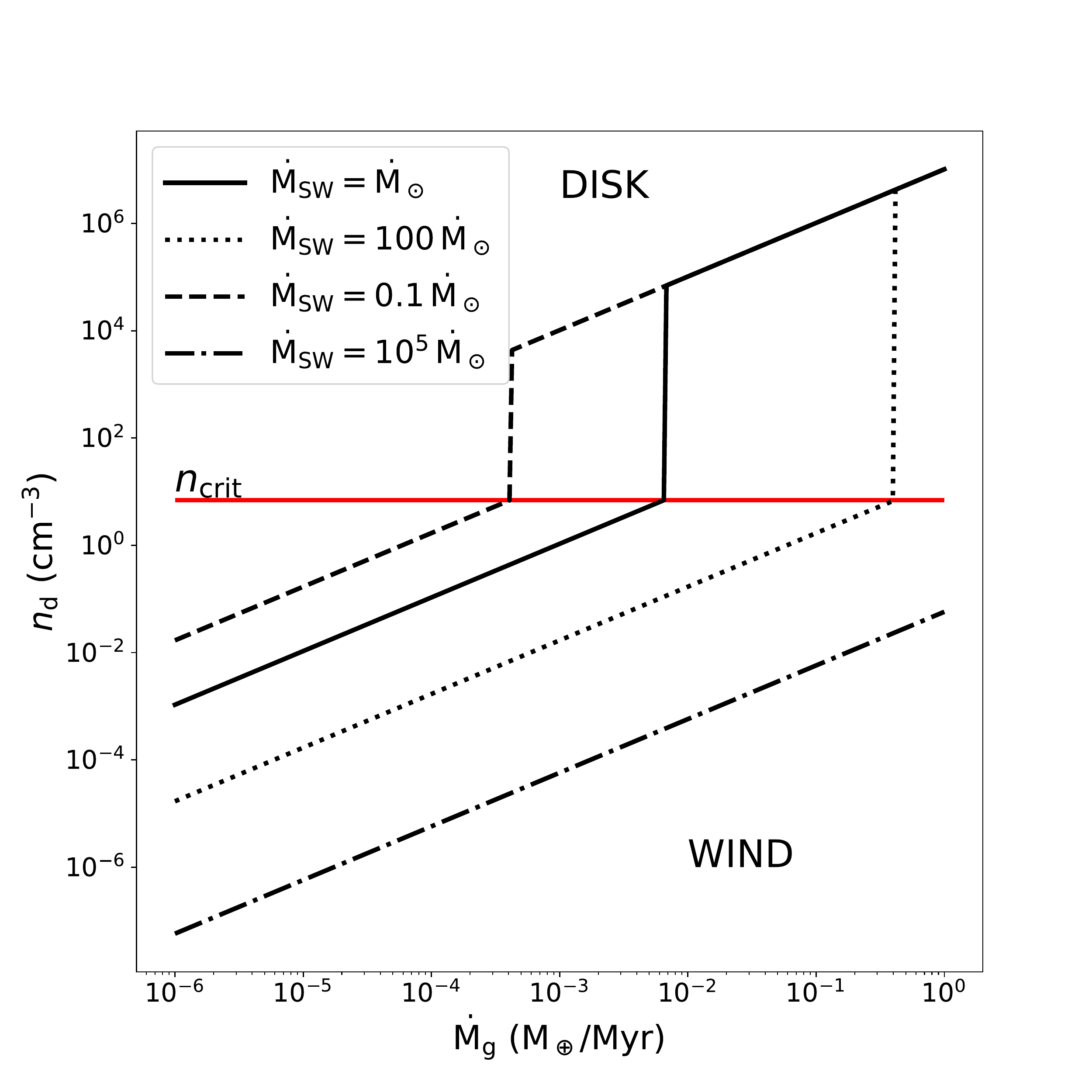}
   \caption{\label{fig0}  Steady-state gas density in the belt (in Keplerian motion) predicted by our model for different values of gas production rate. The line in red represents $n_{\rm crit}$, which is the critical density below which the gas has a wind structure (see Eq.~\ref{ncritneutraleq}) for collisions between SW protons and either C, O, or ions. Above this critical density, the gas is assumed to form a viscous disk with $\alpha=10^{-3}$. The case of high ${\rm{\dot{M}_{\rm SW}}}=10^5$ ${\rm \dot{M}}_\odot$ shows the regime where $t_{\rm leave}/t_{\rm hit} > 1$, independent of ${\rm{\dot{M}_{\rm SW}}}$ where we assumed $v_{\rm SW}=100$ km/s (see main text).}
\end{figure}

\noindent We note that $n_{\rm d}$ is the density of gas in the disk, i.e. with a Keplerian rotation, while the gas density in the wind, moving outwards will be described as $n_{\rm w}$ and their sum by $n_{\rm g}$. The transition between the wind and disk regimes is shown in  Fig.~\ref{fig0} representing the gas disk density as a function of a variety of realistic gas input rate for different stellar mass loss rates. The wind regime lies below the red line representing $n_{\rm crit}$. And because $n_{\rm d}<n_{\rm crit}$ in the regime of interest, we require

\begin{equation}\label{cond2}
\frac{\rm{\dot{M}_g}}{{\rm{\dot{M}_{\rm SW}}}} < \mu \left(\frac{H}{R}\right).
\end{equation}


The derived steady-state density assumes that gas particles leave the disk immediately after impact but in reality it takes a finite time $t_{\rm leave}$. Eq.~\ref{ngasss3} will then break down if the gas particle gets hit by another proton before leaving the belt. To know when this happens we calculate the waiting time before a gas particle gets hit by a proton $t_{\rm hit}$ and compare\footnote{We note that $t_{\rm hit}$ is not linked to the mean free path of protons $\lambda_{\rm w}$ through the disk but rather to the mean free path of a gas particle seeing a flux of incoming protons as described in Appendix~\ref{beltmodel}.} it to $t_{\rm leave}$, which is the time taken by a gas particle to leave the main belt after impact. We find that Eq.~\ref{ngasss3} breaks down when $t_{\rm leave}/t_{\rm hit} \sim  2 \mu \sigma_{\rm col} n_{\rm SW} H  \gg 1$ (see Appendix \ref{beltmodel}). Another condition to use Eq.~\ref{ngasss3} is therefore ${\rm{\dot{M}_{\rm SW}}}<(2 \pi R^2 v_{\rm SW} m_p)/(\sigma_{\rm col} H)$ or in more sensible terms

\begin{equation}\label{cond3}
{\rm{\dot{M}_{\rm SW}}}<3 \times 10^{-10} \, {\rm M}_\odot{\rm /yr}  \left(\frac{R}{100 \, {\rm au}}\right) \left(\frac{v_{\rm SW}}{100 \, {\rm km/s}}\right) \left(\frac{\alpha_X}{\alpha_{\rm CO}}\right)^{-2/3}.
\end{equation}

In the absence of external perturbers, a population of collisionally interacting particles
orbiting a central body always tends to relax toward a dynamical state where relative
velocities are isotropically distributed. In terms of orbital elements this translates into $\langle i \rangle \sim \langle e \rangle/2$, with $i$ and $e$ the orbital inclinations and eccentricities of solids feeding the gas, respectively, so that the scale height of the planetesimals is $\sim R \langle e \rangle/2$ \citep[see][for more details]{2009A&A...505.1269T}. To further simplify our model, we assumed that $H$ is equal to the latter and took a typical $\langle e \rangle \sim 0.1$ value \citep{2009A&A...505.1269T}. We also assumed that $R_{\rm col}$ is in the limit where its radius is fixed by the particle's radius (case valid for collisions with ions or C, or O) but the result would be otherwise 10 times greater if we had considered CO collisions instead. The typical value found in Eq.~\ref{cond3} is much greater than the solar mass loss rate of $\rm{\dot{M}}_\odot=1.4 \times 10^{-14}$ M$_\odot$/yr and should most always be true around $>10$ Myr old stars even for the youngest M-dwarfs such as AU Mic, which wind mass loss rate estimations are in the range $10$-$10^3$ $\rm{\dot{M}}_\odot$ \citep{2009ApJ...698.1068P,2017ApJ...848....4C,2017A&A...607A..65S}.

We note that when Eq.~\ref{cond3} breaks but that Eq.~\ref{cond2} is still valid, it means that each gas particle scatters $t_{\rm leave}/t_{\rm hit} > 1$ wind protons before it is ejected from the disk. As a first approximation this is equivalent to reducing ${\rm{\dot{M}_{\rm SW}}}$ by a factor $t_{\rm hit}/t_{\rm leave}$. The new steady-state density in this case would be roughly equal to $n_{\rm d} (t_{\rm leave}/t_{\rm hit}) = 2 {\rm{\dot{M}_g}} / (4 \pi R \Delta R v_{\rm SW} \mu m_p)$, indeed independent of ${\rm{\dot{M}_{\rm SW}}}$. 

The gas density described by Eq.~\ref{ngasss3} is that of the gas that is not yet kicked out by SW protons and it has a Keplerian velocity. The windy part density can also be derived at the belt location. The number of gas particles hit per unit time is $\rm{\dot{N}_g}$ derived earlier, which can then be used to derive the wind density by dividing it by the velocity of gas particles after impact and the area of solid angle of approximately $\sim \pi^2$ (see Appendix \ref{coloutcome}), which leads to

\begin{equation}\label{nwind}
n_{\rm w}=\frac{{\rm{\dot{M}_g}}}{\pi^2 R^2 m_p v_{\rm SW}},
\end{equation}

\begin{figure}
   \centering
   \includegraphics[width=9.cm]{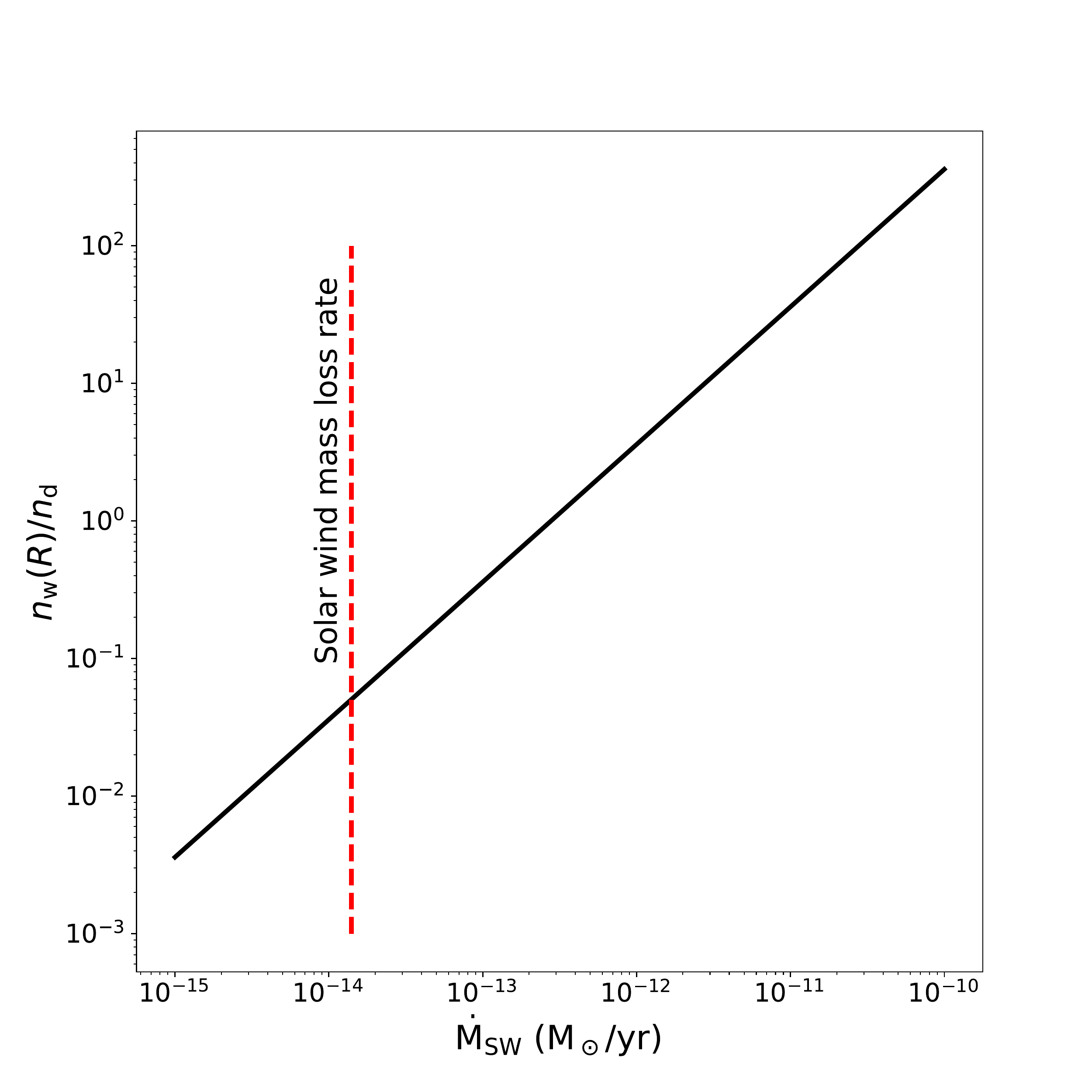}
   \caption{\label{figw} Ratio of density between gas in the wind (spherical outward motion) to that in the belt (in Keplerian motion in the belt) at the belt location $R$ as a function of the stellar wind mass loss rate. For this plot, we use Eq.~\ref{nwtogratio} with $\mu=28$, $v_{\rm SW}=100$ km/s, $R=100$ au, $\Delta R=50$ au, $\sigma_{\rm col}=2 \times 10^{-18}$ m$^2$, and $H=0.05 R$. For a stellar wind mass loss rate of a few tens that of the Sun, values of $n_{\rm w}$ become greater than $n_{\rm d}$ and the wind becomes dominant.}
\end{figure}

\noindent We can therefore work out the ratio of gas in the wind (moving outwards spherically) compared to that in the belt (moving in a Keplerian motion) equal to

\begin{equation}\label{nwtogratio}
\frac{n_{\rm w}}{n_{\rm d}}=\frac{{\rm{\dot{M}}}_{\rm SW} H \Delta R \sigma_{\rm col} \mu}{\pi^2 R^2 m_p v_{\rm SW}},
\end{equation}

\noindent which can become greater than 1 for large values of ${\rm{\dot{M}}}_{\rm SW}$ and the wind would then dominate (see Fig.~\ref{figw}). Typically, for a stellar wind mass loss rate of a few tens that of the Sun \citep[rather typical in young systems,][]{2015A&A...577A..27J,2015A&A...577A..28J}, values of $n_{\rm w}$ become greater than $n_{\rm d}$ and both components are important to consider.

When Eq.~\ref{cond2} breaks then the part of the disk between $R$ and $R+\lambda_w$ would behave as described before while the rest of the disk beyond $R+\lambda_w$ would behave like a standard viscous disk \citep[with the dimensionless $\alpha$ value parametrizing the viscosity,][]{1973A&A....24..337S} as described in the current literature \citep[e.g.][]{2019MNRAS.489.3670K} and in Appendix \ref{gasmodel} (see Eq.~\ref{ngasncriteq}). The viscous spreading happens on rather long timescales (a few Myr to tens of Myr depending on $\alpha$) so that we can neglect its contribution to the inner disk that reaches the previously described steady state on much smaller timescales. In reality the transition may not be as abrupt as shown in Fig.~\ref{fig0} between the two regimes (disk Vs. wind) but we still expect a strong discontinuity because the part of the disk that becomes thick to incoming protons will produce gas particles with outwards velocity of order tens of km/s, which will collide again before leaving the disk and mostly come back to their original velocity after a few collisions. Hence, collisions do not produce escaping gas anymore and gas can accumulate. A numerical model accounting for collisions in a Monte Carlo fashion could describe this transition more accurately but this is left for future work when clear belt wind detections will be at hand.




To sum up, we are left with two gaseous populations after the interaction of the stellar protons with gas released from planetesimals: 1) Gas colocated with the planetesimal belt with a Keplerian velocity, which accumulates before being hit by protons, 2) A spherical wind travelling outwards beyond the belt with a mean velocity of order $v_{\rm SW}/\mu$ ($v_{\rm SW}$ being of order 100-1000 km/s for M to A stars), or typically 5-50 km/s, which leads to a slow moving spherical belt wind. We find that for $\lambda_w \gg \Delta R$ then all released gas is eventually blown out as a wind. Otherwise, only the part between $R$ and $R+\lambda_w$ behaves as a wind and the rest as a standard viscous disk (with $\lambda_w$ quickly becoming very small compared to $R$). For the case of strong ${\rm{\dot{M}_{\rm SW}}}$ when $t_{\rm leave}/t_{\rm hit} > 1$, a wind is still expected for $\lambda_w \gg \Delta R$ but the gas density will be higher than derived in Eq.~\ref{ngasss3} because less particles can be blown out given that one gas particle gets hit by several protons. All these different regimes are shown in Fig.~\ref{fig0}. We note that if there is another belt further out releasing a substantial amount of gas then the outward moving wind could be affected but the details could only be captured by a more complex numerical simulation.




\section{Results}

\subsection{Application of our model to real observations}

\begin{figure}
   \centering
   \includegraphics[width=9.5cm]{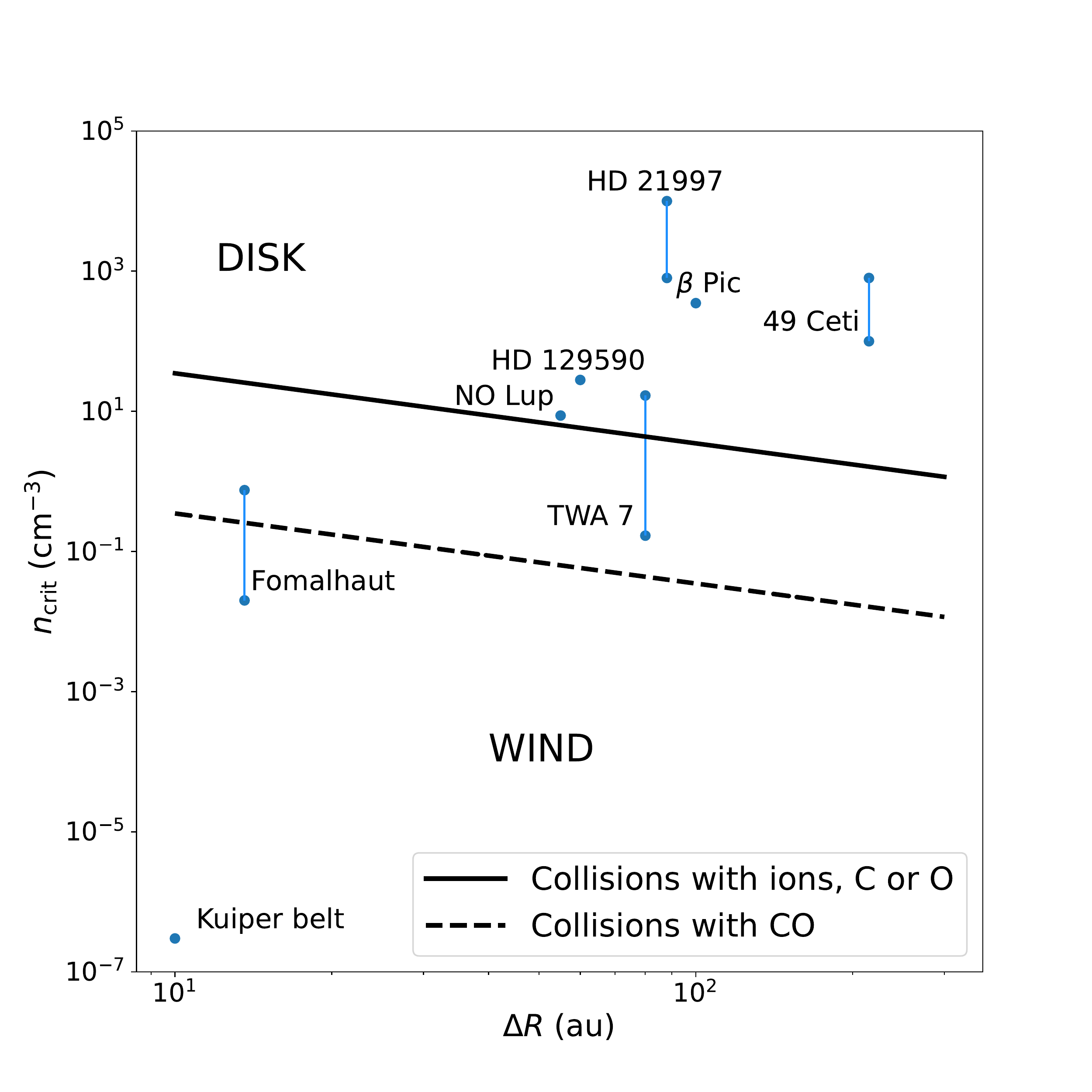}
   \caption{\label{fig1}  The lines represent the critical density $n_{\rm crit}$ at which the gas transition from a disk to a wind Vs belt's width $\Delta R$. The solid line is for a model where $n_{\rm crit}$ is given by Eq.~\ref{ncritneutraleq} and assumes that protons collide with ions, C or O. The dashed line is for collisions between protons and CO. We indicate in capital letters whether the gas would be more DISK- or WIND-like in different regions of the plot. We note that under the black lines, winds are very likely as stated in the main text. The blue points show the gas density $n_{\rm d}$ estimated for the different systems shown in the plot. A vertical line joins two blue dots when a range of values are given in the literature for the gas densities or masses. Data are given in Table~\ref{tab:table1}.}
\end{figure}

We now use our model to compare $n_{\rm d}$ and $\Delta R$ obtained from observations and investigate whether they lie at the top or bottom of the $n_{\rm crit}$ prediction given by Eq.~\ref{ncriteq}. Figure~\ref{fig1} shows $n_{\rm crit}$ as a function of $\Delta R$ where we assumed ions, C, or O (solid) and CO (dashed) colliding with protons. We consider a sample of observed disks with gas (described in Table~\ref{tab:table1}), which includes both high-gas-density systems and low-density ones \citep[such as the KB, for which the gas density is taken from][]{2021A&A...653L..11K}. Above the black solid line, the gas will always have a disk-like structure (except for very early-type stars because of the effect of radiation pressure as explained later). Below the black solid line, a gas wind may form for the ionised, C or O gas species and below the dashed line for neutral CO gas. In Appendix \ref{ionisation}, we show that the ionisation timescale $t_{\rm ion}$ of the gas is in most cases greater than the collisional timescale $t_{\rm hit}$ between neutrals and protons from the SW, so that gas does not ionize fast enough before it leaves the disk (except in bright FUV star systems or with strong winds) and the ionised part will start appearing at larger distances. However, as explained in the previous section, the ionisation fraction of CO is expected to be of order 0.1 because of charge exchanges with SW protons, even at the planetesimal belt location.

In Figure~\ref{fig1}, we notice that the gas in Fomalhaut, TWA 7 and NO Lup (all detected with ALMA) possibly lies in the WIND region of the parameter space given uncertainties. For NO Lup, a wind has recently been detected \citep{2021MNRAS.502L..66L}. For TWA 7 and Fomalhaut, the current gas detections were obtained after integrating in frequency, and cannot rule out a wind with certainty. Deeper images could lead to the first confirmation of a belt wind detection in these systems. TWA 7 being younger may be more favourable but Fomalhaut would give the first possibility to measure the stellar wind properties around an A star. 

ALMA has the power to probe this ``windy'' domain and deep images targeting  systems with low levels of gas could show a wind structure (i.e. gas moving radially outwards at high velocities). For instance, NO Lup is a young K7 class III system with a $\sim$22 km/s outflowing wind detected in CO and it appears to lie right at the edge of the wind/disk transition \citep{2021MNRAS.502L..66L}. Our model could provide the first explanation to this unexpected observation. NO Lup could be in a late T-Tauri stage\footnote{Given the uncertainties on the age of the system, NO Lup could also be a main sequence star of 12-15 Myr \citep[member of UCL,][]{2020AJ....160..186L}.} where stellar mass loss rates would be of order $10^{-10}$ M$_\odot$/yr or slighter larger \citep{1995ApJ...452..736H} explaining why the wind would dominate over the gas in Keplerian motion (i.e. $n_{\rm w}/n_{\rm d}>1$, see Fig.~\ref{figw}) but observations at higher resolution are needed to confirm our hypothesis. The age of NO Lup is not very well constrained because of the unknown membership to either Lupus or Upper Centaurus Lupus, and could be around 1-3 Myr (Lupus), or 12-15 Myr (UCL) \citep{2020AJ....160..186L}. We note that in the T-Tauri case, we expect CO$^+$ to dominate \citep[which is possible if ionisation is faster than the time to escape the disk in young T-Tauri stars,][]{2017A&A...602A.105H} as otherwise a wind would not be present. The observed CO would then only come from recombination, which could be tested with ALMA. For more typical lower stellar mass loss rates if CO impact ionisation is faster than UV photodissociation, and CO$^+$ wind removal is faster than recombination, then we would expect CO$^+$ to be $\sim 0.1$ times as abundant as CO in the wind (see Appendix \ref{ionisation}). Given their similar energy levels, but much stronger transitions for CO$^+$ compared to CO, we expect CO$^+$ J=3-2 to be brighter than CO J=3-2 by a factor of $\sim$ 60 for similar excitation conditions. We therefore indicate CO$^+$ as a potential tracer of disk winds around young main sequence stars.

In the $\beta$ Pic system, atoms such as Ca, Na, Fe are detected at large distances from the star going from $\sim$ 10 to 100s of au \citep{2004A&A...413..681B,2012A&A...544A.134N}. The atomic gas is  in Keplerian rotation, which is surprising because it is expected that radiation pressure should blow out most atomic gas (such as FeI or FeII) on hyperbolic orbits. \citet{2004A&A...413..681B} explain that a braking agent may be at work but that hydrogen does not work. \citet{2006ApJ...643..509F} suggest a few years later that the overabundant ionised carbon may be the braking agent able to explain the observations. In the context of this paper, one may also wonder, whether stellar winds could also be important to push these metals away from the star where they are supposedly produced \citep[e.g. via outgassing exocomets,][]{2014Natur.514..462K}. Using the mapping of Fe I in $\beta$ Pic at the VLT by \citet{2012A&A...544A.134N} we get an estimate of the Fe density at 20 au (their fig. 9) and find $\sim$0.2 cm$^{-3}$. This value being smaller than the critical density found in the paper, we could imagine that indeed Fe or other metals could be pushed outwards due to collisions with stellar protons, in the case where the total metal density (including all species) is not greater than the critical density. However, one would need to look for the specific metal-proton cross sections in their neutral and ionised states to carry out the calculations properly as well as the initial velocity vectors of those metals, which goes beyond the scope of this wind pioneering paper. Qualitatively, we note that when the metals enter the main gas disk in Keplerian rotation \citep[made of CO and carbon,][]{2014Sci...343.1490D,2018ApJ...861...72C} at $>$50 au, the total density may suddenly increase to a value $>$ $n_{\rm crit}$, and the metals are braking naturally by bouncing onto other gas species (in a fluid regime) and are not affected by stellar wind protons anymore. This scenario needs to be studied in more detail and comparisons to observations should be carried out to see if it may be a competing or parallel scenario to that of radiation pressure to explain the presence of metals located at great distances from the star in the $\beta$ Pic system. If realistic, observations of metallic gas winds would be expected in systems with exocomet-like detections (even around stars providing low radiation pressure) and the wind would turn into a stable Keplerian disk if there is a susbtantial gas disk located further away.

\subsection{Detecting a belt wind}
\subsubsection{ALMA}
\begin{figure}
   \centering
   \includegraphics[width=9.cm]{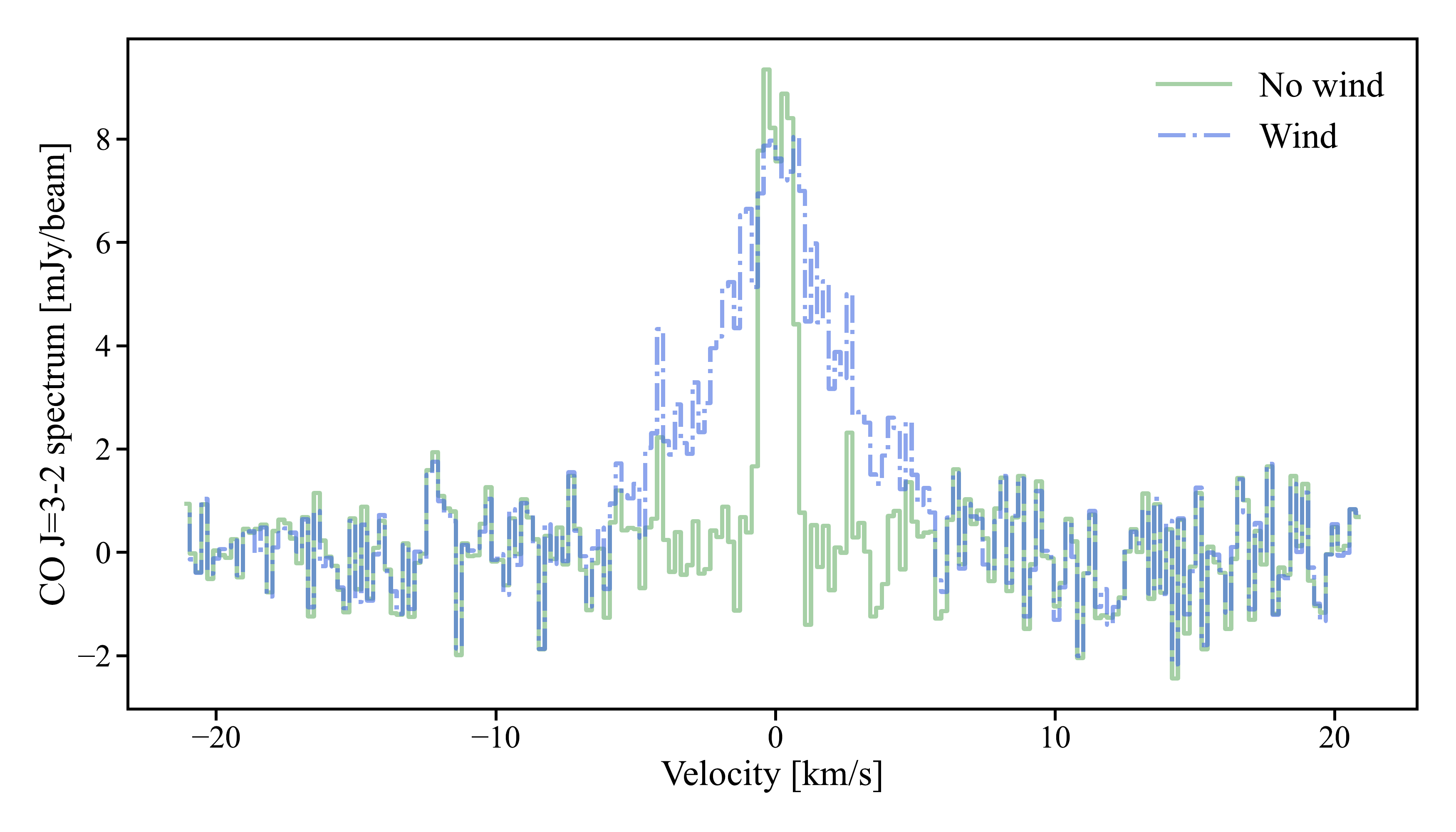}
   \caption{\label{figs} Simulated spectra of the J=3-2 CO line for TWA 7 for the no wind case (solid), and wind (dash-dotted) cases. Along with the moment-0 image of the disk, one can identify that the wind profile is not Keplerian but it is even more clearly visible on the moment-1 images in Fig.~\ref{figm1} (because the velocity structure changes significantly). }
\end{figure}

We now explore in more detail under which conditions ALMA could clearly distinguish between a disk or a wind structure. In Figure~\ref{figs}, we show simulated ALMA spectra for an observation of the gas in TWA 7 for different scenarios: a) a purely Keplerian disk, or b) a wind with a velocity of 20 km/s. The $v_{\rm g}=20$ km/s belt wind would be roughly produced from a stellar wind of velocity $\sim \mu v_{\rm g}$ of $560$ km/s (not atypical for an M-star) assuming a CO-dominated wind. 

For our synthetic observations, we assume that CO has a constant number density radially between 60 and 90 au, and a (radially constant) scale height of 6 au. The velocity field of this component is assumed to be in Keplerian rotation around a star of 0.51 M$_{\odot}$ similar to TWA 7. For the wind component, we assume that its density $n_{\rm w}$ simply follows $n_{\rm w}/n_{\rm d}=0.1$ at the disk location (worst case scenario). In the analytical model we developed, we have not extrapolated our results to large distances. This is because after a collision with a proton, a gas particle can be ejected in any direction along a half sphere which boundary is set by the plane perpendicular to the proton velocity vector, which we call the half sphere of influence. Therefore the emerging wind becomes 3D and it moves spherically outwards, which makes it difficult to follow analytically. Nevertheless, we developed a semi-analytical method to predict what the density and velocity of the wind will look like further away to be able to make accurate predictions for observations (see details in appendix \ref{ALMA}). The results are shown in Fig.~\ref{figwind} representing the (from top to bottom) face-on and edge-on densities, and the velocity structure.

The velocity profile we derive will be used to mimic the expected doppler profile of the gas wind to make synthetic images. As can be clearly seen in Fig.~\ref{figwind}, the gas outflow has a butterfly-like shape in the edge-on direction, with density peaking along the direction of the disk mid-plane. In the head-on direction it logically assumes an isotropic profile that continuously decreases outward.
As for the orientations of mean velocities within the outflow, they schematically radiate from the location of the parent belt, while the absolute magnitude of the velocities is maximum in the prolongation of the belt's mid-plane. This is an expected result as this is the direction for which the kinetic energy transferred by the impacting protons is maximum (see Eq~\ref{eqX}).

\begin{figure}
   \centering
        \includegraphics[width=7.cm]{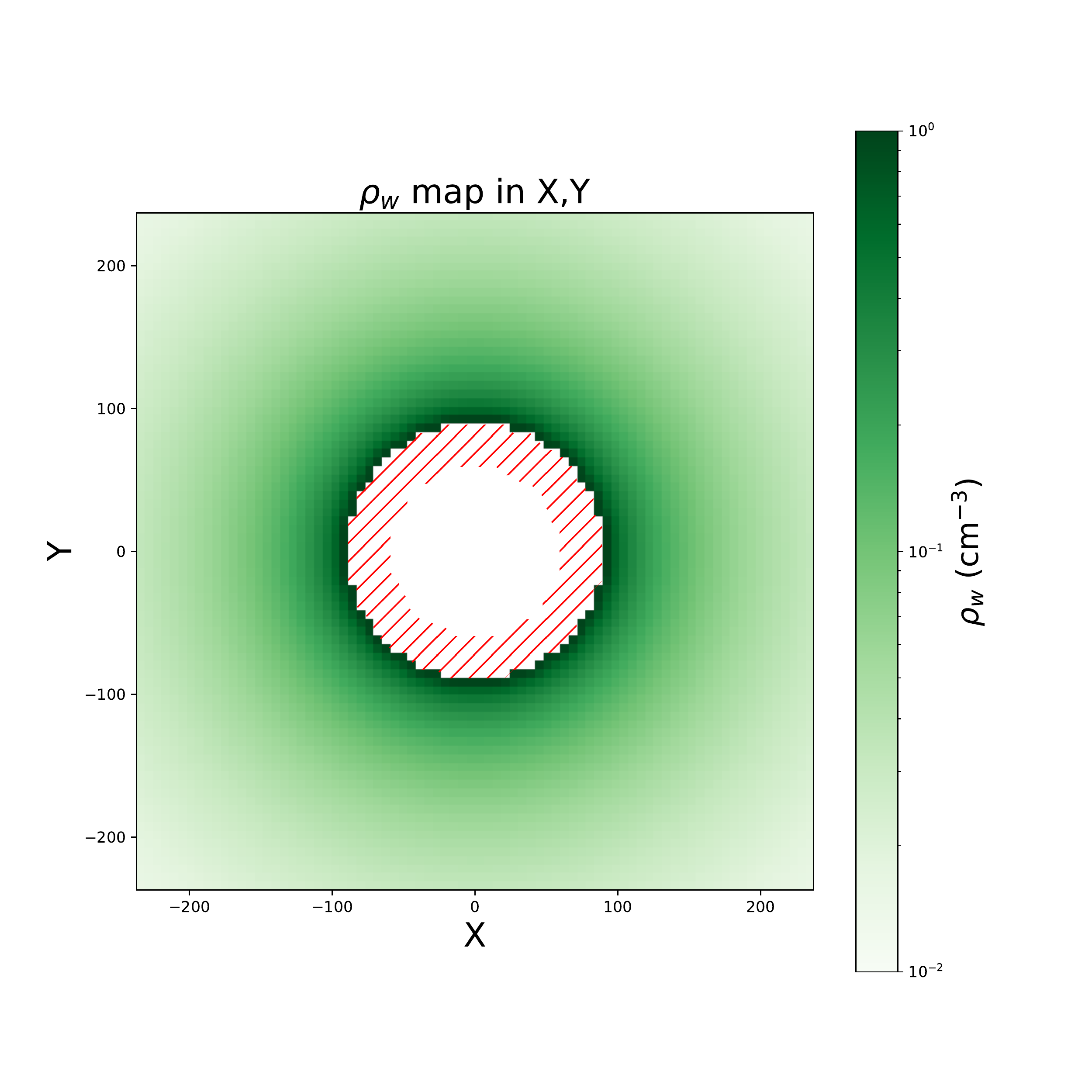}
      \includegraphics[width=7.cm]{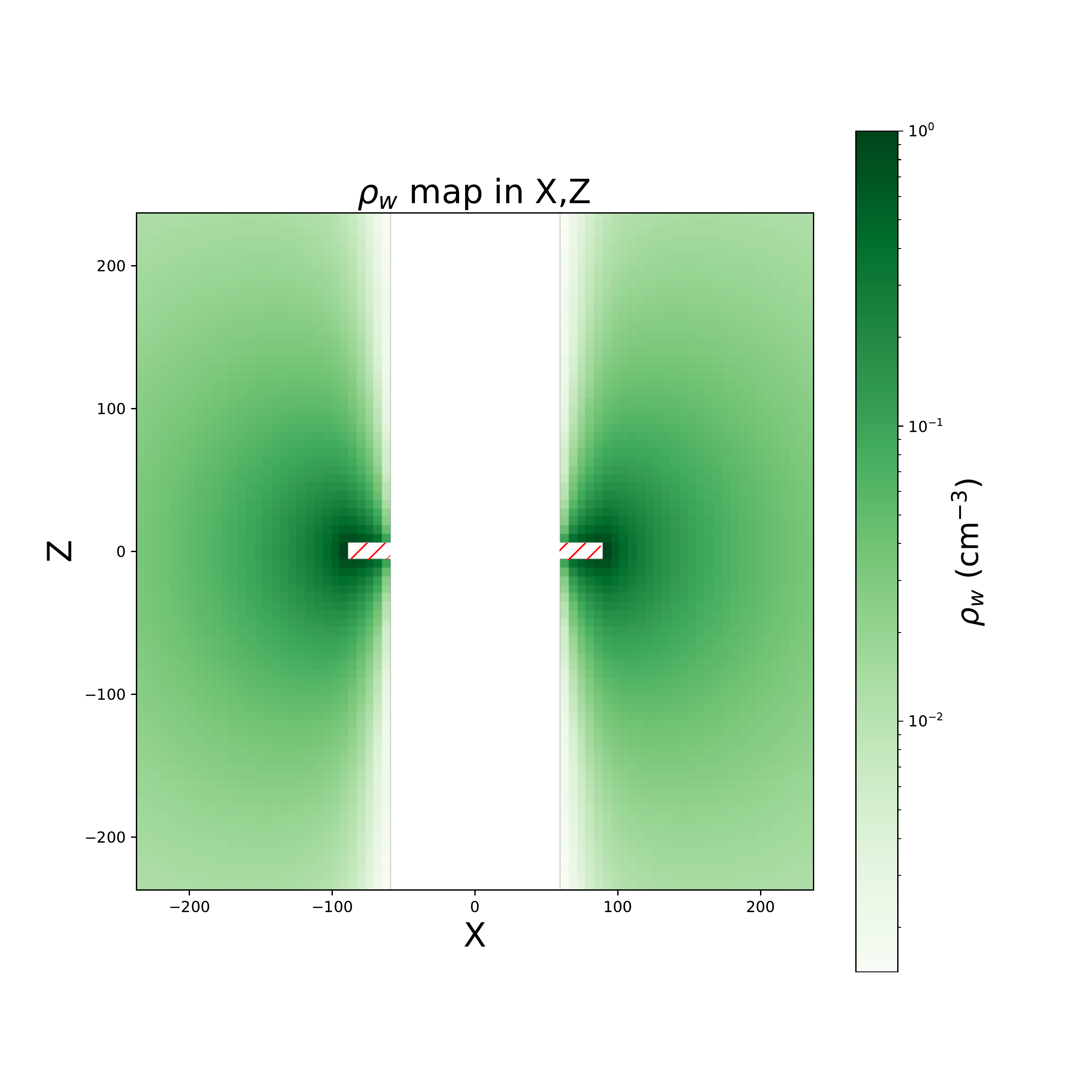}
   \includegraphics[width=7.cm]{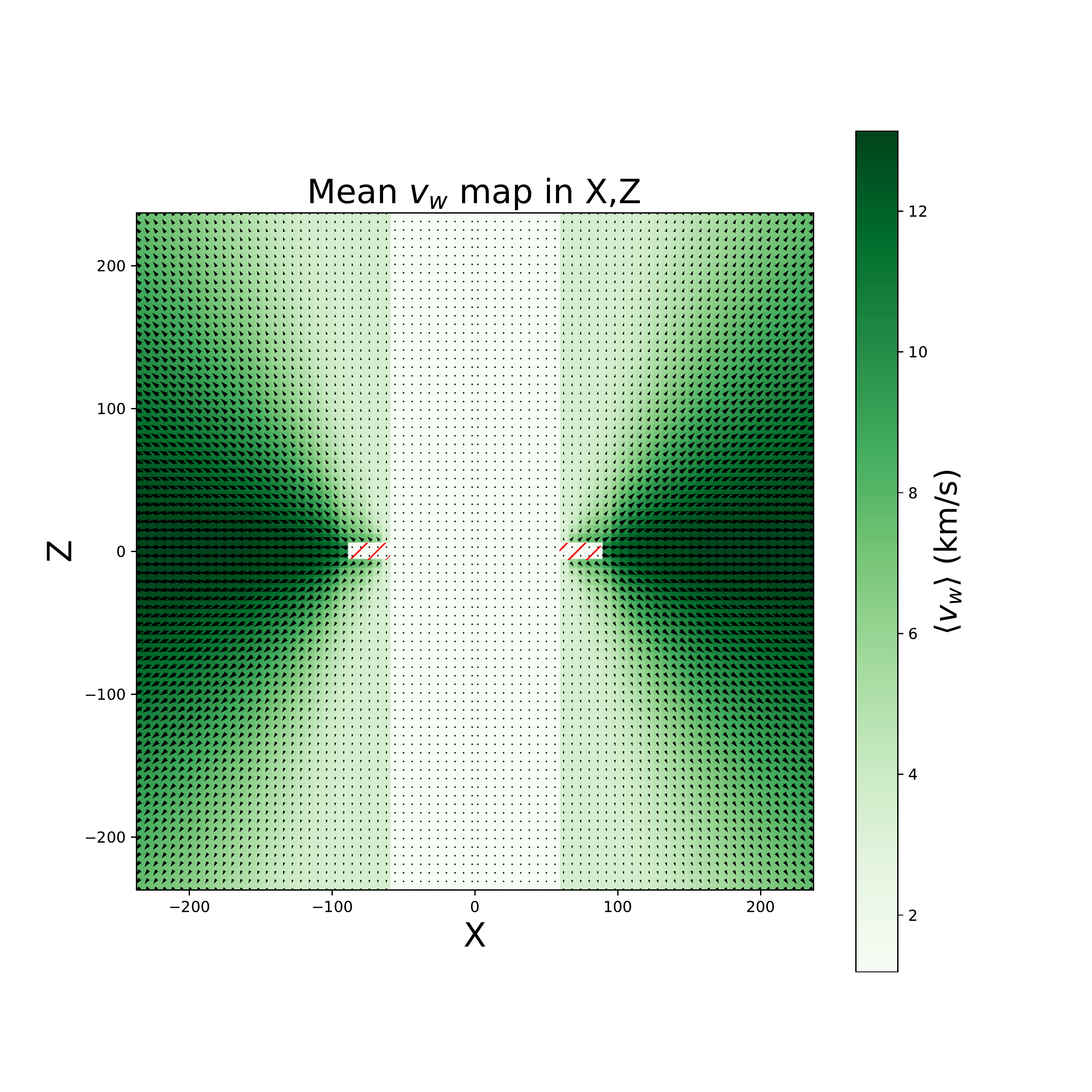}

   \caption{\label{figwind} Wind structure for a belt extending from 60 to 90 au. {\it Top}: Wind density in a (X,Y) plane at Z=0 (cut in the midplane). The density was set to 1 cm$^{-3}$ for the wind emerging from the belt. {\it Middle}: Wind density in a (X,Z) plane at Y=0 (cut at the middle of the belt). {\it Bottom:} Velocity structure (direction and magnitude) integrated along the line of sight (here assumed to be along the Y axis). The red hatches show the location of the belt that releases gas.}
\end{figure}

\begin{figure}
   \centering
   \includegraphics[width=9.cm]{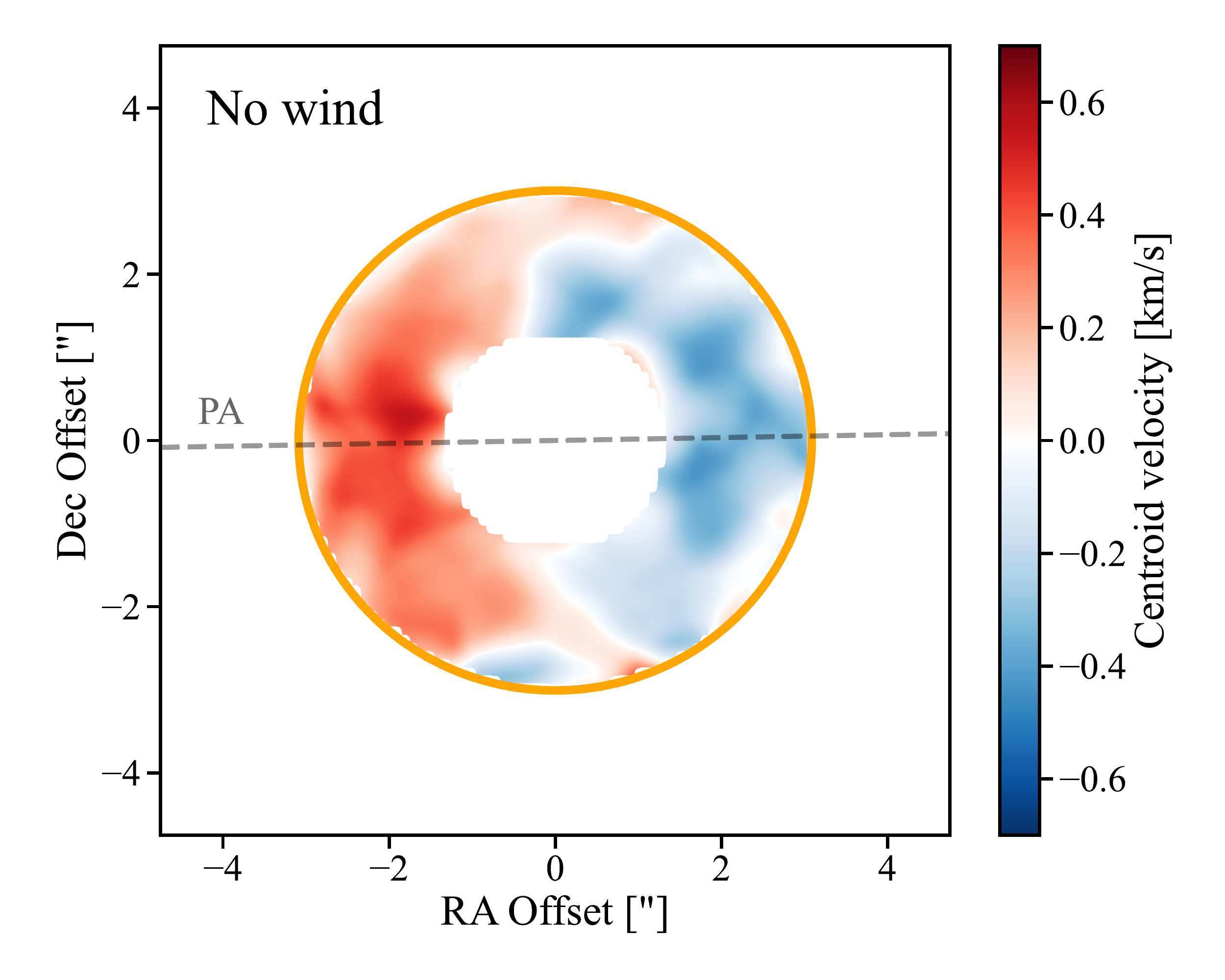}
      \includegraphics[width=9.cm]{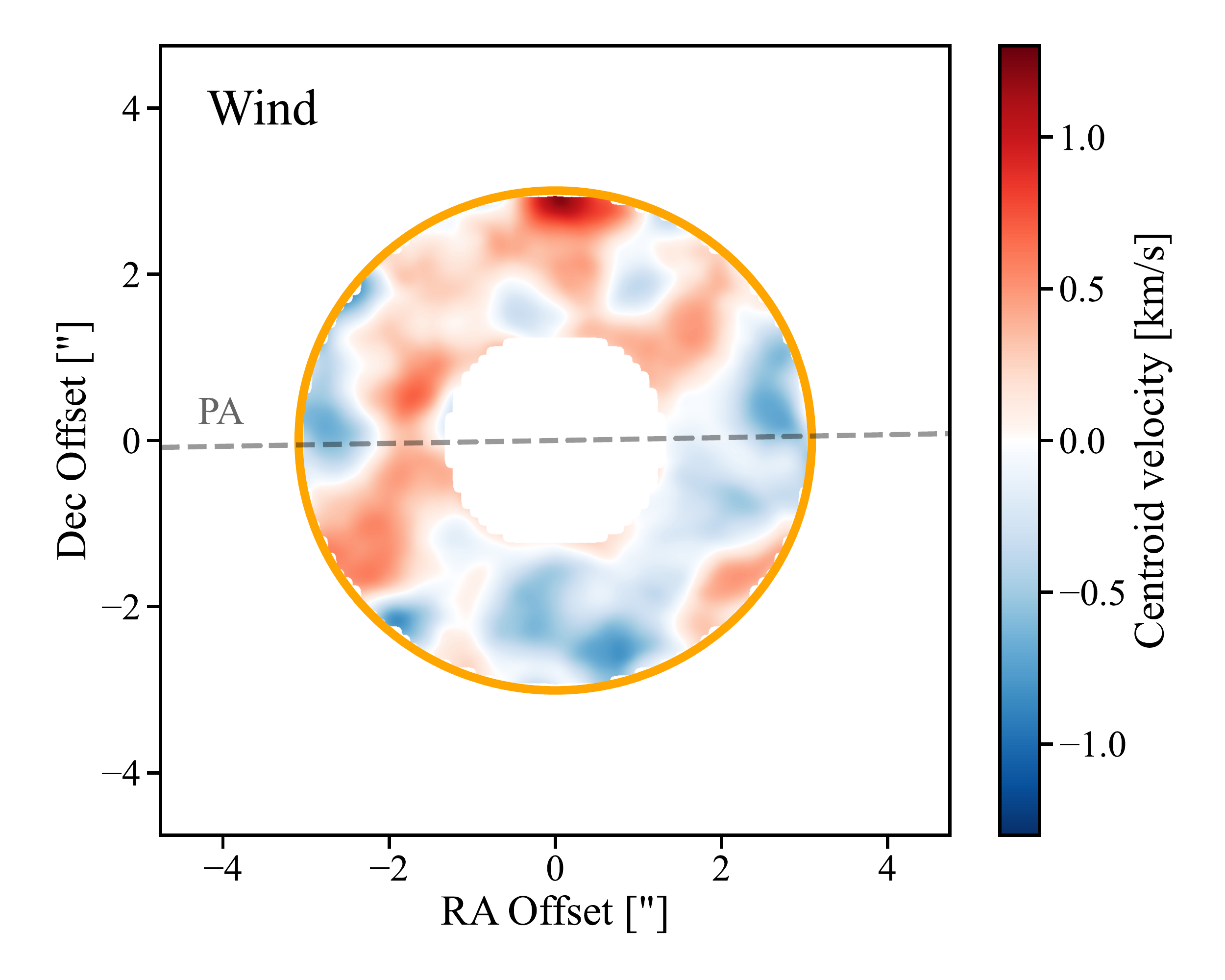}

   \caption{\label{figm1} Moment-1 images (intensity-weighted velocity) for the no wind (top) and wind (bottom) cases. The no-wind Keplerian case is aligned on the disk position angle while it is very inclined or even perpendicular for the wind case, making them easily distinguishable, which can also be seen on the position-velocity diagram (Fig.~\ref{figpv}). }
\end{figure}

\begin{figure}
   \centering
   \includegraphics[width=9.cm]{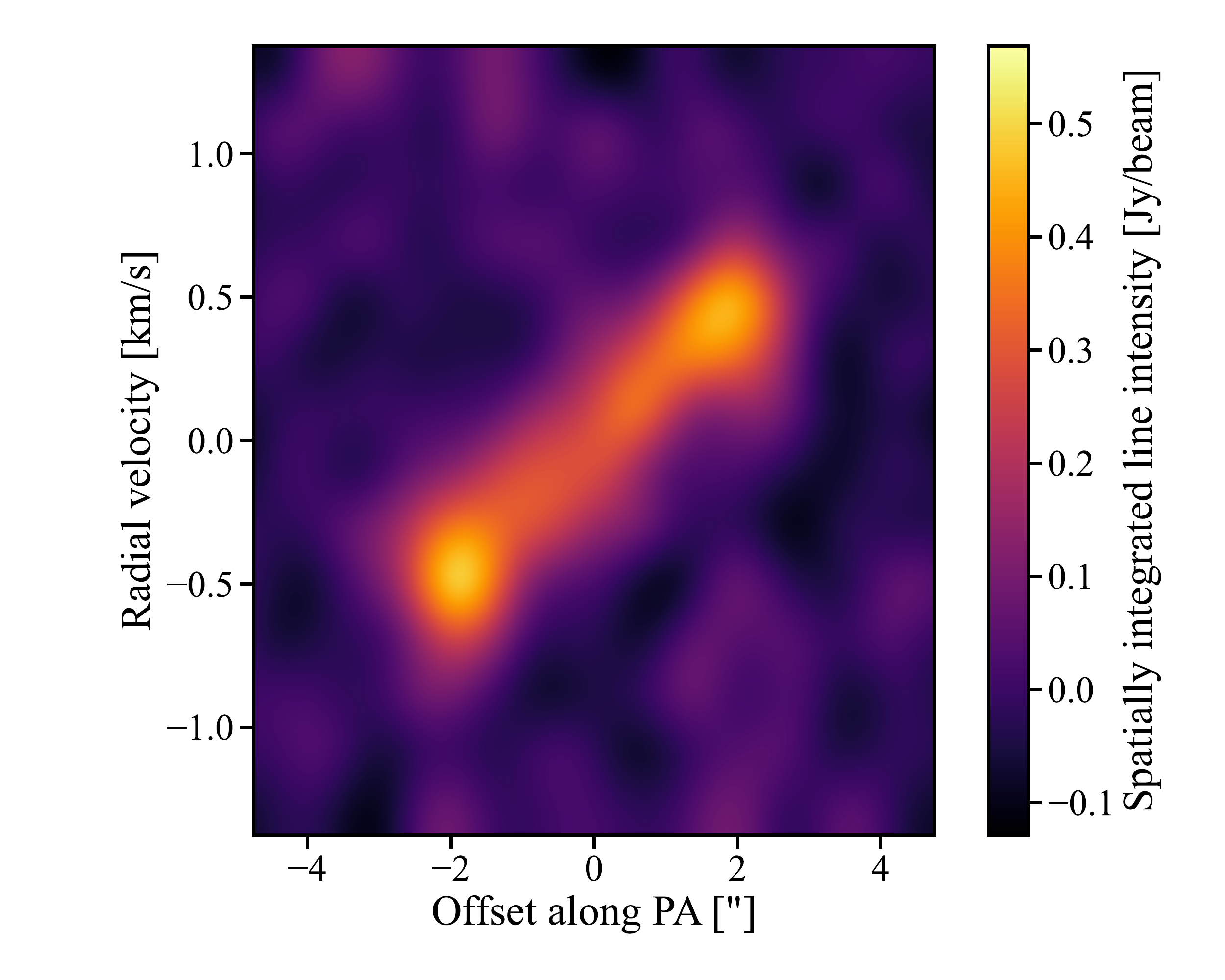}
      \includegraphics[width=9.cm]{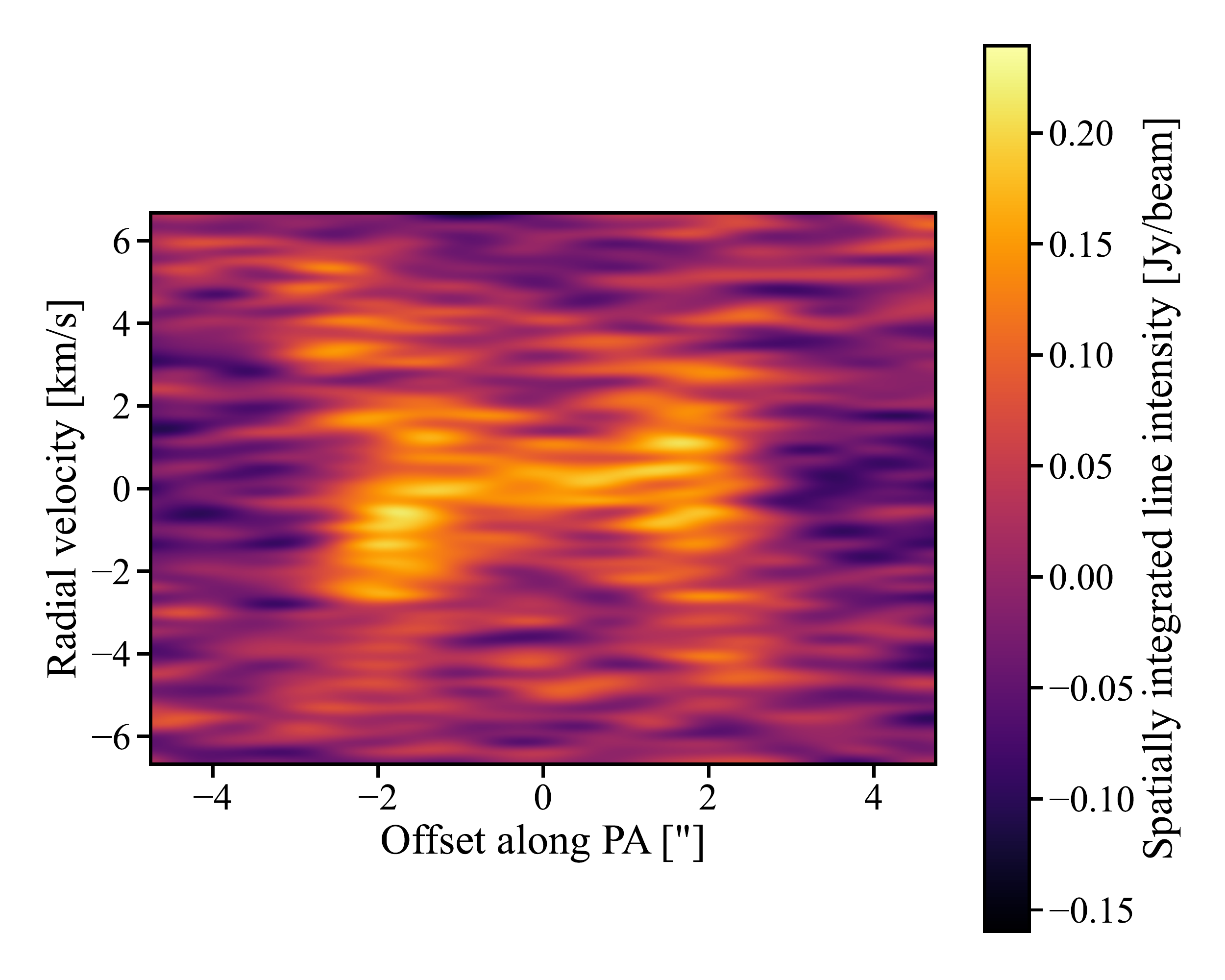}

   \caption{\label{figpv} Position-Velocity (PV) diagrams for the no wind (top) and wind (bottom) cases. The no-wind Keplerian case shows a typical PV diagram for a Keplerian disk along its position angle while it is horizontal for the wind case, making them easily distinguishable. }
\end{figure}

For building the synthetic images, we also assume a $\sim$ 1" resolution corresponding to ALMA in its most compact (C-1) configuration, which gives a $\sim$40 au resolution at TWA7's distance. We find that we can distinguish between a purely Keplerian disk model and a wind model for a sensitivity of 1.5 mJy/beam for a spectral resolution of 244.14 kHz (corresponding to a velocity resolution of 0.21 km/s), which requires roughly 5 hours on source with 43 antennas. The 20 km/s wind can clearly be distinguished because of its wider non-Keplerian spectrum when compared to the gas position given by moment-0 images (see Fig.~\ref{figm0}). Moreover, in Fig.~\ref{figm1}, we show the moment-1 velocities (intensity-weighted velocity) for the different scenarios and find that the wind case can clearly be distinguished from the no-wind case given their very different radial velocity structures. In the case of the 20 km/s wind, the velocity is perpendicular to the position angle of the disk \citep[that is obtained thanks to optical observations for the case of TWA 7,][]{2018A&A...617A.109O}, while it is aligned with it for the no-wind case, which is also clear on the position-velocity diagrams (see Fig.~\ref{figpv}).

We thus conclude that from these types of observations, we could clearly distinguish ``belt'' winds with ALMA and use them to access to the SW velocities and densities around main-sequence stars, which is otherwise difficult \citep{2015A&A...577A..27J}. 

Indeed, from the detection and using Eq.~\ref{ngasss3} with $n_{\rm dobs}$ (the observed disk density), we can go back to the factor ${\rm \dot{M}_g}/{\rm \dot{M}}_{\rm SW}$.
If the wind is detected then we can use the simple one zone model presented in the paper to get some first estimates of the stellar wind velocity assuming that $v_{\rm SW} \approx \mu \langle v_{\rm wobs} \rangle$ (the mean value of the observed wind velocity derived from the redshift of the wind compared to the star velocity and after correcting for potential orientation effects). We can also retrieve a value of ${\rm \dot{M}}_{\rm SW}$, which does not depend on ${\rm \dot{M}_g}$ (hence independent of a gas release model) using Eq.~\ref{nwtogratio}, i.e. by computing the ratio between $n_{\rm wobs}$ (the observed wind density) and $n_{\rm dobs}$. In appendix \ref{velanddens}, we describe the process in more detail.

\subsubsection{In the UV or optical?}
Observations of those winds may also be possible in the UV, as implied by the recent detections of a gas wind in $\eta$ Tel \citep{2021AJ....162..235Y}. However, the feasibility of far-UV absorption studies from space relies on two factors. The first is the viewing geometry of the wind, which needs to cross the line of sight and produce a sufficiently high column density for detection. As shown in this paper, the column density would be highest through the disk midplane, therefore favouring an edge-on viewing geometry. But we note that if the wind density is high compared to that of the disk (in Keplerian rotation), there is more leeway on the inclination because even for a face-on geometry, some gas particles will still cross our line of sight since they can be ejected perpendicular to the midplane. The second factor is the presence of a strong, detectable background, which is typically the central star. However, the star's flux density at the relevant UV wavelengths \citep[$\sim$1500\AA\, for the main CO A-X bands, see e.g.][]{2000ApJ...538..904R} is on the Wien side of the stellar emission, and therefore very strongly dependent on the spectral type, distance from Earth, and the presence of UV excess, if any. As the viewing geometry and stellar flux density are very system-specific, it would be difficult to make a prediction on whether winds could be generally detectable at UV wavelengths or not.
A more promising and widely-applicable avenue could be detection of CO$^+$ in the near-UV/optical (in addition to the CO$^+$ lines that can be targeted with ALMA), as is common for Solar System comets, in the violet region of the optical \citep[$\sim$4200\AA\, A-X bands, e.g.][]{2018ApJ...854L..10C}. Absorption studies in this range would benefit from a much improved strength of the stellar continuum, allowing a wider range of spectral types and distances from Earth to become accessible. The CO$^+$ bands are however complex, and would require a specific line list with adequate codes to handle the radiative transfer. We therefore deem a thorough exploration of a CO$^+$ detection in an exoplanetary system in this wavelength range to be beyond the scope of this paper.

\subsection{Predictions of debris disk systems with belt winds}

\begin{figure}
   \centering
   \includegraphics[width=9.cm]{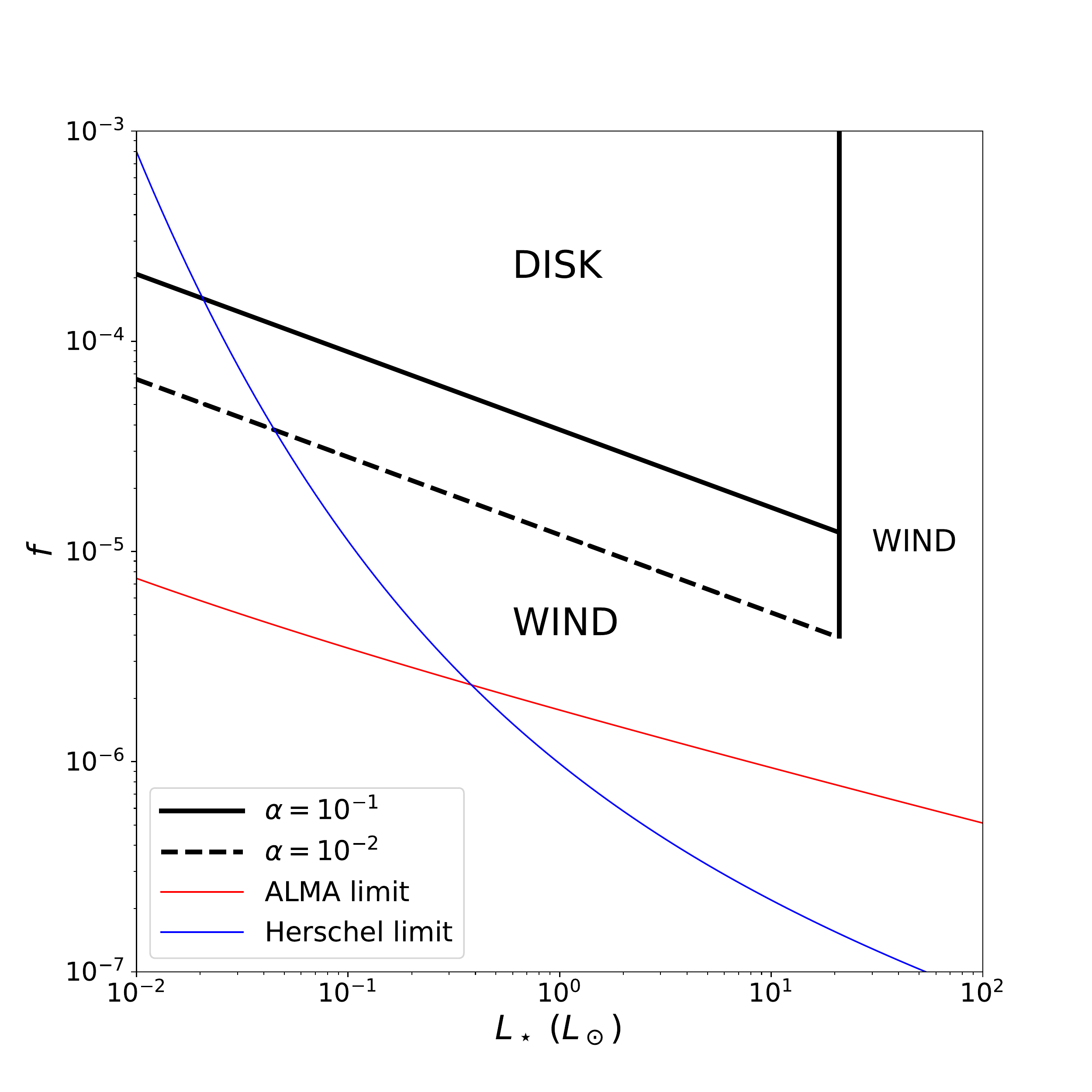}
   \caption{\label{figf} Transition between disk and wind regions shown in the space of the fractional luminosity of a debris disk as a function of the host star luminosity. The bottom left and far right regions are likely to be populated by gas having a wind structure while the upper left region would more likely be filled by circumstellar gas disks. We show the limit between disk and wind cases for two values of $\alpha$: 0.1 (solid) and 0.01 (dashed). The thin red and blue lines are dust detection limits of ALMA at 870 microns and Herschel at 70 microns assuming dust lies at 100 au. See the assumptions behind this plot at the end of the Appendix~\ref{gasmodel}.}
\end{figure}

Finally, we assess how to predict a gas wind presence from dust observations of debris disks only (which are much more numerous than those of gas) and more specifically from their fractional luminosity (the infrared luminosity over that of the star) and stellar luminosity, which are accessible for hundreds of debris disks (see details in Appendix \ref{gasmodel}). To do so, we assume that gas is injected in the belt at a rate $\rm{\dot{M}_g}$ and then evolves viscously with an $\alpha$ prescription \citep{2016MNRAS.461..845K,2016MNRAS.461.1614K}. We assume that the gas disk reaches a steady-state and compare the gas density to $n_{\rm crit}$ to find when the gas density becomes so low that gas starts behaving as a wind. It leads to Eq.~\ref{finaleq2}, which can be further simplified to
 
 \begin{equation}\label{feq}
f<10^{-5} \left( \frac{L_\star}{L_\odot} \right)^{-0.37} \left( \frac{\Delta R}{50 \, {\rm au}} \right)^{-1/2} \left(\frac{\alpha}{10^{-2}}\right)^{1/2},
\end{equation}
 
 \noindent where $f$ is the fractional luminosity of a debris disk and $L_\star$ the host star luminosity. We use this equation and the fiducial values used in it to make Fig.~\ref{figf}, which shows the DISK Vs. WIND regions for varying $f$ and $L_\star$. Therefore, a wind-like structure is expected for disks with low fractional luminosities, of order $10^{-5}-10^{-4}$ for M stars and down to $10^{-6}-10^{-5}$ for the more massive A-stars. It shows that late type stars are expected to create gas winds more readily than, e.g., A-type stars. 
 
However, for A stars with too high luminosities $L_\star \gtrsim 20 L_\odot$ a gas wind is also expected because of the radiation pressure becoming too high on the gas as was shown in Fig.~11 of \citet{2017MNRAS.469..521K}, and more recently by \citet{2021AJ....162..235Y} with a criterion on the temperature such that $T_\star>10,200$ K. We note that when the gas starts becoming optically thick to UV photons, this mechanism would stop working \citep{2017MNRAS.469..521K}, which needs to be computed for each observation. Indeed, \citet{2017MNRAS.469..521K} predict that in all systems with $\dot{M}_{\rm CO} > 10^{-4}$ M$_\oplus$ Myr$^{-1}$, ionised and neutral carbon can be protected from being blown out thanks to shielding. Without shielding, ionised carbon would
be blown out for systems with $L_\star >15 L_\odot$ but neutral oxygen would stay bound
up to 25 $L_\odot$. This region of the parameter space with low fractional luminosities is expected to be the most populated by population synthesis models of debris disks for A stars \citep{2007ApJ...663..365W} and for F, G, K stars \citep{2018MNRAS.475.3046S}. We note that Eq.~\ref{feq} is giving an order of magnitude because the gas density derived from the standard gas disk model we use may not be an exact good fit of observations for a specific system (and $\alpha$ could vary). Thus, debris disks with slightly larger fractional luminosities such as TWA 7 or Fomalhaut can still be in the WIND regime because gas observations show that they indeed have low gas levels.
  
\section{Conclusions}
In this paper, we explore whether the gas that has been detected in many debris disk systems may be blown out as a wind due to collisions with high-velocity protons from stellar wind, rather than being a circumstellar disk in Keplerian rotation as assumed in current models. We find that, indeed, this wind-like behaviour of the gas might be present in systems in which the gas density is low, typically below the threshold value of $7 \, (\Delta R/50 {\rm \, au})^{-1}$ cm$^{-3}$, where $\Delta R$ is the belt's width. We find that the analytical model we developed may explain the outflowing gas detection in the young NO Lup system. Moreover, the two systems with the lowest gas masses detected to date, TWA 7 and Fomalhaut, may already be in this wind region and further observations with ALMA could clearly disentangle between a disk or a wind structure. More generally, we find that debris disks with low fractional luminosities such that $f \lesssim  10^{-5} (L_\star/L_\odot)^{-0.37} $ are naturally expected to form ``belt'' winds. Likewise, gas is expected to have a wind-like structure around early type stars with luminosities $\gtrsim 20 \, L_\odot$ because of the action of radiation pressure rather than stellar wind (see cartoon in Fig.~\ref{figcartoon} illustrating all the different gas dynamics expected for varying gas densities and stellar luminosities). In addition, we argue that gas released very close to the central star by, e.g., FEB-like bodies (falling evoparating bodies or star-grazing exocomets) could also be affected by stellar winds and may explain some observations in the $\beta$ Pic system or lead to further interesting detections around systems with detected FEBs. We find that CO$^+$ may be a good tracer of belt winds created because of the action of stellar winds and it can be targeted with ALMA and probably in the optical. Future detections of belt winds would allow us to retrieve the SW properties (mass loss rate and velocity) around a large variety of main-sequence stars, which is otherwise difficult to measure, especially around A stars where no measurements have led to a detection so far.
   
\begin{acknowledgements}
This paper is dedicated to Florian. We thank the referee for a helpful review that improved the quality of the paper. QK thanks Hervé Beust and Alex de Koeter for interesting discussions about induced dipole interactions and stellar mass loss rates, respectively. QK also thanks Alexandre Faure and Evelyne Roueff for discussions about elastic cross sections.
\end{acknowledgements}


\begin{appendix}

\section{Stellar wind model}\label{stellarwindmodel}

In this paper we are interested in stellar winds produced around both young and old stars, for spectral types going all the way from A to M (O and B stars are not considered here). 
Stellar wind measurements are scarce ($\sim$ 12 systems) around main-sequence stars because they are difficult to detect \citep{2011ApJ...741...54C}. For A stars, there are no detection so far and the mechanism behind their winds still needs to be assessed firmly \citep{1992A&A...257..663L,2014A&A...564A..70K}.

There are two fundamental properties for SWs: their velocity $v_{\rm SW}$, and density $n_{\rm SW}$ (that is linked to the stellar mass loss rate $\rm{\dot{M}_\star}$).
To assess the values of those properties, we use models such as that of \citet{2015A&A...577A..27J} that describe SW properties for dwarf main-sequence stars (M, K, G, F type stars). This model is an extension of a state-of-the-art Solar wind model. The mechanisms that heat the solar corona and accelerate the solar wind remains unknown to this day. The two main competing hypotheses are that the solar wind could be driven by the dissipation of Alfvén waves and turbulence, or by magnetic reconnection events \citep[e.g.,][]{2010ApJ...720..824C}. In a similar way, there remains many uncertainties about winds around main-sequence A stars that cannot be probed easily. Theoretically, no significant winds are expected for stars with spectral types later than B6 for which the line driven mass loss rate should be smaller than $10^{-12}$ M$_\odot$/yr \citep{2014A&A...564A..70K}. An empirical upper limit is reported in \citet{1992A&A...257..663L} leading to a less constraining value of $1-2 \times 10^{-10}$ M$_\odot$/yr. There could be other mechanisms than line driven winds driving stellar mass loss in A stars (e.g. magnetism, pulsation, coronal heating) but only observations could tell us more. If none of these mechanisms can drive winds efficiently in A stars, then the mass loss rate could even be smaller than for our Sun, i.e. $<1.4 \times 10^{-14}$ M$_\odot$/yr. Winds around O- and earlier B-stars are more powerful and are expected to lead to higher velocities and densities but they are not considered in this study because the number of debris disks found around these stars are only a handful. Winds around younger T-Tauri or Herbig stars are also expected to be larger, reaching values greater than $10^{-10}$ M$_\odot$/yr \citep{1995ApJ...452..736H,1995A&A...302..169N}.

As it is noted in these models, there are large uncertainties on the properties of the SWs around stars of A to M spectral types, and we will explore a large range of values to account for that. The wind velocities can roughly vary from 100 to 1000 km/s (it is close to the value of the escape velocity at the star surface) and the star mass loss rates on the main sequence can go from 0.1 to 1000 that of our Solar System where $\rm{\dot{M}}_\odot=1.4 \times 10^{-14}$ M$_\odot$/yr, and larger for even younger stars on the pre-main sequence. For instance, for the emblematic AU Mic system, theoretical estimates for the stellar mass loss rate varies from 10 \citep{2009ApJ...698.1068P} to 1000 $\rm{\dot{M}}_\odot$ \citep{2017ApJ...848....4C,2017A&A...607A..65S}.

From the stellar mass loss rate, we can then find the SW density used in this paper through the relation that follows

\begin{equation}\label{nwindfrommdot}
n_{\rm SW}=\frac{\rm{\dot{{M}}_\star}}{4 \pi R^2 v_{\rm SW} \mu_w m_p},
\end{equation}

\noindent with $\mu_w=0.6$, the mean molecular weight of the wind, based on the Solar wind \citep{2015A&A...577A..27J} and we use the notation ${\rm{\dot{{M}}_{\rm SW}}}$ (instead of ${\rm{\dot{{M}}_\star}}$) when only accounting for protons from the stellar wind in the main text and then take $\mu_w=1$ in the previous equation to get the stellar wind proton density.

We note that the star mass loss rate depends on the star radius and mass, and stellar angular velocity $\Omega_\star$, and scales as $\Omega^{1.33}_\star {\rm R}^2_\star {\rm M}^{-3.36}_\star$ \citep{2015A&A...577A..27J} so that the Solar wind was predicted to be an order of magnitude stronger in its youth and more generally, young systems are expected to have stronger SWs because of the faster rotation of the central star \citep{2015A&A...577A..28J}. The angular velocity of main-sequence stars can vary but is highest between 10-100 Myr and can reach 100 $\Omega_\odot$ with a mean of $\sim 10 \, \Omega_\odot$ for solar type stars and low-mass stars \citep{2013EAS....62..143B}. As a first approximation, the angular velocity scales with time as $t^{-1/2}$. Our model also accounts for saturation effects, i.e. there is a limit to which mass loss rates can increase due to an increase of $\Omega_\star$, which happens at $\Omega_{\rm sat}=15 \Omega_\odot (M_\star/M_\odot)^{2.3}$, where $\Omega_\odot=2.67 \times 10^{-6}$ rad/s is the Carrington rotation rate. We use this SW model in Appendix \ref{ionisation} to work out typical SW collisional timescales.


\section{Belt wind model}\label{beltmodel}
Here, we describe an idealised belt wind model for a planetesimal belt comprised between $R$ and $R+\Delta R$ with a scale height $H$ and a constant density throughout. The gas producing the belt wind is released from planetesimals \citep[e.g., by collisions or sublimation,][]{2021A&A...653L..11K} and then pushed outwards by impacting stellar wind protons.

Let us first describe the regime where the rate of gas production is small relative to the stellar wind, i.e. it is the case described in \citet{2021A&A...653L..11K} for the Kuiper belt, where all gas particles get hit and ejected outwards by the stellar wind, thus creating a belt wind.

Let us define a wind proton mass loss rate $\rm{\dot{M}_{\rm SW}}$ and a gas production rate in the disk $\rm{\dot{M}_g}$. We first assume that the mean free path of wind protons in the gas $\lambda_w=1/(\sigma_{\rm col} n_{\rm g})$ is much greater than the belt's width $\Delta R$, where $\sigma_{\rm col}$ is the proton/gas particle collision cross-section. 

For instance, in the case of ion collisions (e.g. C$^+$ colliding with protons), we expect the elastic cross-section to be small because charged particles are repulsive at long range. To quantify that, we use that the kinetic energy of the proton equals its electrostatic energy to get

\begin{equation}\label{Rcol1}
R_{\rm col}=\frac{e^2}{2 \pi \epsilon_0 m_p v_{\rm col}^2}
\end{equation}

\noindent where $e$ is the elementary charge, $\epsilon_0$ the vacuum permittivity, $m_p$ the proton mass, and $v_{\rm col}$ the relative velocity between ions and protons. We find that for typical velocities $>100$ km/s, $R_{\rm col}$ becomes smaller than the typical radius of ions so that we should use the radii of ions instead. We thus take $R_{\rm col} \sim R_X$, where $R_X$ is the typical radius of the particle modelled by an imaginary hard sphere (Van der Waals radius) given by $(3 \alpha_X/(4 \pi))^{1/3}$. We take $\alpha_{\rm CO}=1.953$ \AA$^3$, $\alpha_{\rm C}=1.760$ \AA$^3$, and $\alpha_{\rm O}=0.802$ \AA$^3$ $\,$\citep{1978AdAMP..13....1M,1997CP....223...59O} and will use $\alpha_{\rm CO}$ in our fiducial model. This leads to $R_{\rm CO} \sim 0.78$ \AA \, or a collision cross-section of $\sim 2 \times 10^{-20}$ m$^2$.

For collisions between protons and CO, we use that the elastic cross-section with protons at 30-100 eV is $\sim 2 \times 10^{-18}$ m$^2$ \citep{1992CP553,2006JChPh.124c4314D}. Typically, $R_{\rm col} \sim 8$ \AA. 

For collisions between protons and neutral C or O atoms (which have no permanent dipoles like CO), we use that the proton induces a dipole on the neutral atom so that \citep{1989A&A...223..304B,2007A&A...466..201B}

\begin{equation}\label{Rcol2}
R_{\rm col}=\left(\frac{e^2 \alpha_X}{\pi \epsilon_0 m_{\rm red} v_{\rm col}^2}\right)^{1/4}
\end{equation}

\noindent where $\alpha_X$ is the polarisability of species $X$, $v_{\rm col}$ the relative velocity between protons and neutrals, and $m_{\rm red}$ is the reduced mass of the two colliders approximately equal to $m_{\rm p}$, the proton mass, when a proton collides with a more massive neutral. When $R_{\rm col}$ becomes smaller than the actual radius of the particle, we use the latter instead. For a typical wind speed greater than 100 km/s, we thus take $R_{\rm col} \sim R_X$, where $R_X$ is the typical radius of the particle. This leads to $R_{\rm CO} \sim 0.78$ \AA \, or a collision cross-section of $\sim 2 \times 10^{-20}$ m$^2$ similar to that of ions.

In the case that $\lambda_w \gg \Delta R$, the reasoning in the main text led to the conclusion that the gas density at steady state in this regime is

\begin{equation}\label{ngasssz}
n_{\rm d}=\left(\frac{\rm{\dot{M}_g}}{{\rm{\dot{M}_{\rm SW}}}}\right) \left(\frac{R}{\mu H \Delta R \sigma_{\rm col}}\right),
\end{equation}

\noindent and the assumption we made that $\lambda_w \gg \Delta R $, together with Eq.~\ref{ngasssz} leads to

\begin{equation}\label{cond}
\frac{\rm{\dot{M}_g}}{{\rm{\dot{M}_{\rm SW}}}} \ll \mu \left(\frac{H}{R}\right),
 \end{equation}

\noindent so that when the proton wind rate becomes too small or the gas production rate too high, the steady state gas density calculated above needs to be changed (see later). To give an idea of the contribution of the wind with a spherical outward motion compared to gas in the belt in Keplerian motion, we plot $n_{\rm w}/n_{\rm d}$ as a function of ${\rm{\dot{M}_{\rm SW}}}$ in Fig.~\ref{figw}. We note that the wind profile (density and velocity) at large distances from the belt is also computed in Appendix \ref{ALMA} to be able to make predictions for future ALMA observations.

When computing the steady-state above, we have assumed that once a gas particle was hit by a stellar proton, it left the disk immediately, which may become untrue if the time to leave the disk after impact $t_{\rm leave}$ becomes much longer than
the time $t_{\rm hit}$ before a gas particle gets hit by a proton once it is released from a planetesimal. We calculate that $t_{\rm hit}=1/(n_{\rm SW} \sigma_{\rm col} v_{\rm SW})$. The gas particle and the proton typically undergo a large angle collision and the post-collision velocity is roughly isometric (see Appendix \ref{coloutcome}) so that most particles will travel less than $2 H$ before leaving the disk. The mean speed of the gas particle after the collision with a proton is given by $v_g \approx v_{\rm SW} / \mu$. We may thus estimate $t_{\rm leave} = \mu (2 H/v_{\rm SW})$ and get

\begin{equation}\label{tleavethit}
\frac{t_{\rm leave}}{t_{\rm hit}} \sim 2 \mu \sigma_{\rm col} n_{\rm SW} H.
\end{equation}

\noindent Eq.~\ref{ngasssz} is valid as long as $t_{\rm leave}/{t_{\rm hit}}<1$ and breaks down when this ratio becomes $>1$, which corresponds to ${\rm{\dot{M}_{\rm SW}}}>(2 \pi R^2 v_{\rm SW} m_p)/(\sigma_{\rm col} H)$. In this case, and if we also have ${\rm{\dot{M}_{\rm SW}}}>({\rm \dot{M}_g}/\mu)(R/H)$ to account for Eq.~\ref{cond}, we expect that the particles are hit multiple times by different protons before leaving the disk thus reducing the wind efficiency to eject gas particles (e.g. two protons only eject one gas particle instead of two). The gas density that accumulates between $R$ and $R+\Delta R$ will be $n_{\rm d} (t_{\rm leave}/t_{\rm hit})$, the fraction in parenthesis accounting for the reduced efficiency of the proton wind to blow gas particles out. Thus, we find in this regime 

\begin{equation}\label{ngasss2}
n_{\rm d}=\frac{2 {\rm{\dot{M}_g}}}{4 \pi R \Delta R v_{\rm SW} \mu m_p},
\end{equation}
  
 \noindent which is independent of ${\rm{\dot{M}_{\rm SW}}}$.

\begin{figure}
   \centering
   \includegraphics[width=9.5cm]{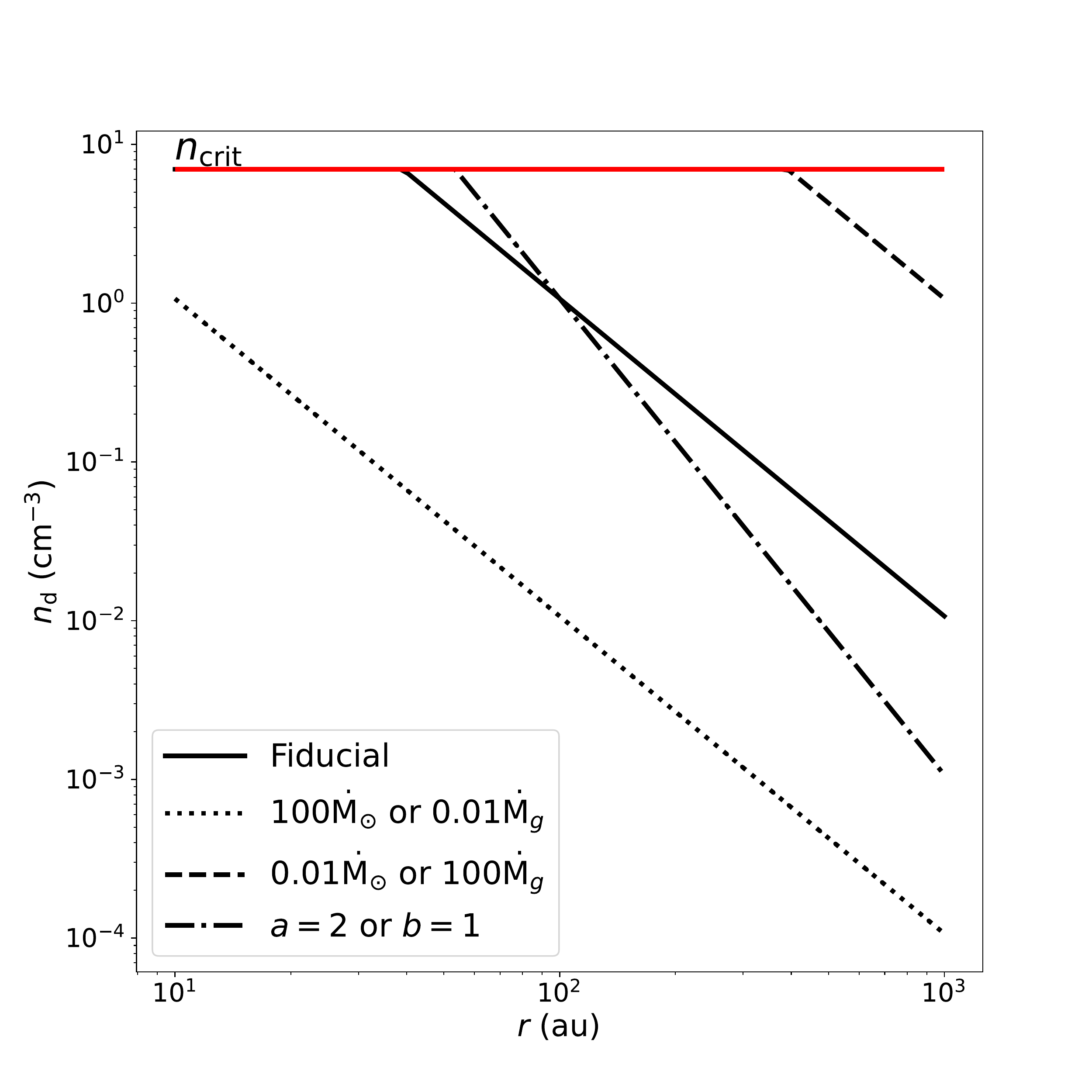}
   \caption{\label{fig6}  Steady-state gas density in the belt predicted by our model as a function of $r$, the radial distance to the star. The fiducial model has $a=1$, $b=0$, $R_0=100$ au, $h_0=0.05$, ${\rm \dot{M}_{\rm SW}}=1.4 \times 10^{-14}$ M$_\odot$/yr, and ${\rm \dot{M}_g}=10^{-3}$ M$_\oplus$/Myr (see Eq.~\ref{ngasss}). Values below $R_{\rm crit}$ (see Eq.~\ref{rcrit}), i.e., above the red line, are not shown. The density beyond the planetesimal width $\Delta R$ will mostly scale as $r^{-2}$ as the fiducial model (see Appendix \ref{ALMA}) but it is not represented in the plot.}
\end{figure}

If we now assume that the gas surface density is not constant but given by $\Sigma = \Sigma_0 (r/R_0)^{-a}$ and $H/r=h_0 (r/R_0)^b$, then following the same procedure as described before to find the steady-state but using d$r$ annuli instead, we find that $n_{\rm d}=\Sigma/(2 H \mu m_p)$ can be defined as a function of its radial distance $r$ as follows

\begin{equation}\label{ngasss}
n_{\rm d}(r)=\frac{\Sigma_0}{2 \mu m_p R_0 h_0} \left(\frac{r}{R_0}\right)^{-a-b-1},
\end{equation}

\noindent where $\Sigma_0=({\rm{\dot{M}_g}}/{\rm{\dot{M}_{\rm SW}}}) (2 m_p/\sigma_{\rm col})(R_0/\Delta R)$. We plot $n_{\rm d}(r)$ in Fig.~\ref{fig6} for various values of a, b, ${\rm{\dot{M}_{\rm SW}}}$, and ${\rm{\dot{M}_{\rm g}}}$. We note that in this case, there is a minimum radius $R_{\rm min}$ below which $n_{\rm d}(r)$ can become greater than $n_{\rm crit}$ (the critical gas density below which a gas wind forms, see Eq.~\ref{ncriteq}) and in this case, the gas would become optically thick to protons and the gas structure would be disk-like. For this not to happen, $R_{\rm min}$ should remain greater than 

\begin{equation}\label{rcrit}
R_{\rm crit} \sim R_0 \left( \mu h_0 \left(\frac{{\rm{\dot{M}_g}}}{{\rm{\dot{M}_{\rm SW}}}} \right)^{-1} \right)^{1/(-a-b-1)},
\end{equation}

\noindent assuming that $\Delta R \lesssim R_{\rm min}$, otherwise it gets more thorough to work out. We also note that beyond the planetesimal belt of width $\Delta R$, the gas wind density will mostly scale as $(r/R)^{-2}$ (see Appendix \ref{ALMA}) assuming that the rate of collisions with stellar protons becomes negligible.

\section{Gas model and comparison to $n_{\rm crit}$}\label{gasmodel}

Here, we describe the standard gas disk model we use to compare to $n_{\rm crit}$ and find the range of debris disk global parameters that will populate the DISK Vs WIND regions in Fig~\ref{figf}. For a disk of scale height $H$ and mean molecular weight $\mu$, the gas density can be linked to the gas surface density $\Sigma_g(r)$ as follows

\begin{equation}\label{ngasncriteq}
n_{\rm g}=\frac{\Sigma_g}{2 H \mu m_p},
\end{equation}

\noindent and the surface density for a disk located at $R_0$ where gas is input at a rate ${\rm \dot{M}_g}$ is equal to  \citep{2019MNRAS.489.3670K}

\begin{equation}\label{sigmaeq}
\Sigma_g(r)=\frac{{\rm \dot{M}_g}}{3 \pi \nu_0}
\left\{
    \begin{array}{ll}
        \left(\frac{r}{R_0}\right)^{\gamma - 3/2} &\mbox{for } r<R_0 \\
       \left(\frac{r}{R_0}\right)^{\gamma - 2} & \mbox{for } r>R_0
    \end{array}
\right.,
\end{equation}

\noindent where $\nu_0=\alpha c^2_s(R_0)/\Omega(R_0)$ is the viscosity at $R_0$ and the temperature scales as $T \propto R^{-\gamma}$. We note that in this case, $n_{\rm g}$ scales as $R^{3\gamma/2-3}$ for $r<R_0$, and as $R^{3\gamma/2-7/2}$ for $r>R_0$ at steady state.

Now, we want to find when $n_{\rm g}=n_{\rm crit}$ to be able to see which part of the parameter space will fall in the DISK Vs WIND regions. As a first approximation, we compute the gas density at $R_0$ to compare to our one zone model critical density (Eq.~\ref{ncriteq}) and after rearranging the equality, we find 

\begin{equation}\label{reseq}
{\rm \dot{M}_g}=\frac{6 \pi \alpha \mu m_p}{\sigma_{\rm col} \Delta R} \frac{c^3_s(R_0)}{\Omega^2_0},
\end{equation}

\noindent which gives the order of magnitude gas production rate where there is a change of regime from a disk to a wind structure.

To relate the debris disk properties (e.g. luminosity, radius) to the previous equation, we use the fact that the gas input rate is related to the mass loss rate of the belt \citep{2017MNRAS.469..521K} through

\begin{equation}\label{masslossrateeq}
{\rm \dot{M}_g}=\delta f^2 \left(\frac{L_\star}{L_\odot}\right)^{13/12} \left(\frac{R_0}{{\rm 1 au}}\right)^{-1/3} {\rm M}_\oplus {\rm Myr}^{-1},
\end{equation}

\noindent where $\delta=\zeta e^{5/3} (2700/\rho)/(2.4 \times 10^{-10} {\rm d}r /r {Q^\star_D}^{5/6})$, with $\zeta$ the ratio of gas-to-dust mass loss rates assumed to be of order 10\%, $f$ the fractional luminosity of the debris disk, $L_\star$ the stellar luminosity (in $L_\odot$), $e$ the mean eccentricity of the parent belt planetesimals, ${\rm d}r$ the belt width (in au), $Q^\star_D$ the collisional strength of solids (in J kg$^{-1}$) and $\rho$ their bulk density (in kg/m$^3$).
Using the same typical values as in \citet{2017MNRAS.469..521K} for $e$, ${\rm d}r$, $Q^\star_D$ and $\rho$, we find $\delta=2.9 \times 10^{4}$.

As we are interested in the limit between DISKS and WINDS for neutral or ion collisions at high velocity ($>$100 km/s), we use that $\sigma_{\rm col}=\pi R^2_{\rm col}$ with $R_{\rm col}$ is set to the CO radius (case for collisions between CO$^+$, C, or O, and protons). We then find that the condition for planetesimal-generated gas to assume a disk-like structure reads

\begin{multline}\label{finaleq}
0.6 \left(\frac{\alpha}{10^{-2}}\right)^{-1} \left( \frac{R_0}{100 \, {\rm au}}\right)^{-10/3} \left( \frac{f}{10^{-5}}\right)^{2} \left( \frac{T_0}{30 \, {\rm K}}\right)^{-3/2} \\ \left( \frac{\mu}{28}\right)^{1/2} \left( \frac{L_\star}{L_\odot}\right)^{1.37}  \left( \frac{\Delta R}{50 \, {\rm au}} \right) > 1,
\end{multline}

\noindent where we assumed $L_\star \propto M^{3.5}_\star$. 

Moreover, if we assume the empiric law derived by \citet{2018ApJ...859...72M} showing that the radial location of a debris disk peak density varies with stellar luminosity following $R_0=73 {\rm au} \, (L_\star/L_\odot)^{0.19}$, then Eq.~\ref{finaleq} simplifies to

\begin{multline}\label{finaleq2}
0.7 \left(\frac{\alpha}{10^{-2}}\right)^{-1} \left( \frac{f}{10^{-5}}\right)^{2} \left( \frac{T_0}{30 \, {\rm K}}\right)^{-3/2} \\ \left( \frac{\mu}{28}\right)^{1/2} \left( \frac{L_\star}{L_\odot}\right)^{0.74}  \left( \frac{\Delta R}{50 \, {\rm au}} \right) > 1,
\end{multline}

For gas in debris disks, $\alpha$ values could be very high (of order 0.1) because of the high ionisation fraction in these disks that may give birth to a strong magneto-rotational instability \citep[MRI,][]{2016MNRAS.461.1614K}. Indeed, fitting of the current gas observations favour high $\alpha$ values of order 0.1 as shown in \citet{2019MNRAS.489.3670K,2020MNRAS.492.4409M}. However, smaller values could also be realistic because non-ideal MRI effects may start being important at low gas densities (e.g. $\alpha=10^{-4}$) and observations of such disks in the future may allow to probe that more clearly. 

\section{Collision outcome between a wind proton and a gas particle}\label{coloutcome}

An elastic collision between a proton of initial velocity $\vec{v_p}$ and position $\vec{r_p}$ with a gas species $X$ of initial velocity $\vec{v_X}$ and position $\vec{r_X}$ will lead to a post-collision velocity $\vec{v'_X}$ for species $X$ given by

\begin{equation}\label{eqX}
\vec{v'_X}=\vec{v_X}-\frac{2}{\mu+1}\frac{\langle \vec{v_X} - \vec{v_p} | \vec{r_X}-\vec{r_p}  \rangle}{\lVert \vec{r_X}-\vec{r_p} \rVert^2}(\vec{r_X}-\vec{r_p})
\end{equation}

\noindent and for the proton

\begin{equation}\label{eqp}
\vec{v'_p}=\vec{v_p}-\frac{2\mu}{\mu+1}\frac{\langle \vec{v_p} - \vec{v_X} | \vec{r_p}-\vec{r_X}  \rangle}{\lVert \vec{r_p}-\vec{r_X} \rVert^2}(\vec{r_p}-\vec{r_X}),
\end{equation}

\noindent where $\mu$ is the mean molecular weight of the gas particle being hit by the proton.

\begin{figure}
   \centering
   \includegraphics[width=8.cm]{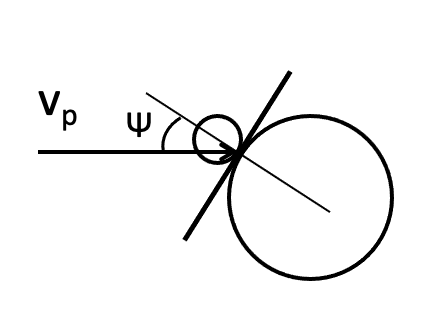}
   \caption{\label{figpsi} Schematic of the collision between a proton of initial velocity $\vec{v}_p$ with a gas particle. The collision angle $\psi$ is between the normal to the surfaces of the spheres at the point of contact and the proton velocity.}
\end{figure}

Given that the proton velocity is much higher than that of the gas particle, and further assuming that the proton moves radially with $\lVert \vec{v_p} \rVert = v_{\rm SW}$, the equation for $\vec{v'_X}$ can be simplified to $2 v_{\rm SW} \cos(\psi) / (\mu+1)$, where $\psi$ is the collision angle (shown in Fig.~\ref{figpsi}). Assuming that the two particles are two spherical balls, the collision angle is roughly given by the angle between the proton velocity and the normal to the surfaces of balls at the point of contact ($\psi=0$ for a head on collision and $\pi/2$ for edge-on) and $\vec{v'_X}$ is parallel to the normal to the surfaces of balls at the point of contact. The angle $\psi$ can vary between [0\,,\,$\pi/2$] and after a collision with a proton, the gas particle most likely exits the disk via its vertical height rather than crossing its whole belt of width $\Delta R>H$. We note that there will be less collisions with $\phi$ angles close to $\pi/2$ because high impact angles cover less impact parameters that are uniformly distributed. The distribution of impact angles scales as $\cos(\phi)$, meaning that the impact angles are well represented by a cone pointing radially outwards with an opening angle $\phi_m$ of $\sim 50$ deg (because almost 4/5 of collisions will be in that cone). We deduce that the molecules escape into a solid angle of approximately $2 \pi \times 2 \phi_m \sim \pi^2$.

\section{Ionisation timescales}\label{ionisation}

Carbon can be ionised by energetic ($>$11.26eV) photons from the ISM on a timescale of roughly 120 yr \citep{2009A&A...503..323V}, which corresponds to a photoionisation rate of $3 \times 10^{-10}$ s$^{-1}$. Carbon, oxygen and CO can also be ionised by stellar radiation, as is the case in our Solar System where the photoionisation rate is $5 \times 10^{-7} (1 \, {\rm au}/r)^2$. Ionisation of O and CO can also happen through charge exchange with SW protons with a cross-section of $\sigma_{\rm exc} \sim 1.3 \times 10^{-19}$ m$^{2}$ \citep{1997A&A...317..193I,2017PhPro..90..391L}, which translates into an ionisation rate of $1/t_{\rm ion}=n_{\rm SW} \sigma_{\rm exc} v_{\rm SW}$, where $t_{\rm ion}$ is the ionisation timescale (see appendix \ref{stellarwindmodel} for the details of the SW model we use). This is roughly an order of magnitude smaller than the collisional timescale for CO meaning that some CO$^+$ may appear even in CO dominated-regions with a factor one to ten in density. We also use spectra from \citet{2003IAUS..210P.A20C} to get typical stellar fluxes $F$ (in erg/s/cm$^2$/nm) for different stellar types and ionisation cross-sections $\sigma_i$ (in cm$^{-2}$) from \citet{2017A&A...602A.105H} to compute the photoionisation rates $p_X$ (in s$^{-1}$) for a species $X$ such that $p_X=\int^\infty_0 \sigma_i(\lambda) F(\lambda)/(h \nu(\lambda)) \, {\rm d}\lambda$ with $\lambda$ the wavelength in nm and $h\nu$ in erg, where most cross-section is in the UV. 

The ionisation timescale should be compared to the collisional timescale between protons and neutrals, because after one collision with a fast SW proton ($\gtrsim$ 100 km/s), the gas particle becomes unbound and leaves the system very quickly, i.e. it rapidly reaches outer regions \citep{2021A&A...653L..11K}. The collisional timescale between SW protons and neutrals is $t_{\rm hit}=1/(n_{\rm SW} \sigma_{\rm col} v_{\rm SW})$. 

After a collision, the gas particle direction is isometric (inside of the cone that points outwards of the collision direction) and will usually leave the disk at high angles (see Appendix \ref{coloutcome}), thus travelling roughly $2 H$ before escaping the belt rather than crossing its whole length $\Delta R$. We need to see how the time to leave the disk $t_{\rm leave}$ compares to $t_{\rm ion}$. The timescale for a gas particle to travel $2H$ of the disk given its post-collision velocity (see Appendix \ref{beltmodel}) is roughly $2H\mu/v_{\rm SW}$. It is important to calculate $t_{\rm leave}$ to understand whether gas particles get ionised before or after leaving the disk.

\begin{figure}
   \centering
   \includegraphics[width=9.5cm]{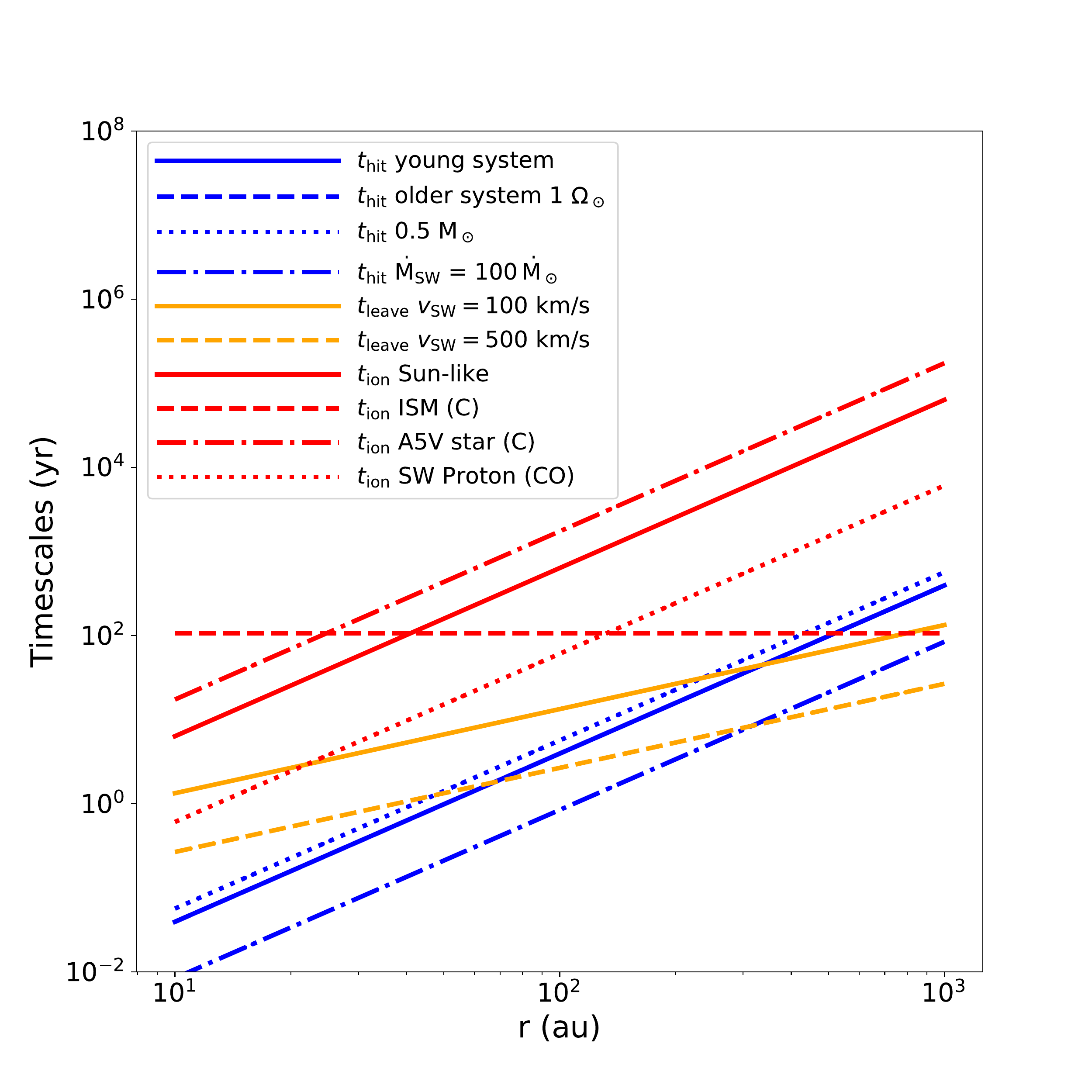}
   \caption{\label{fig2} Timescales for ionisation $t_{\rm ion}$ (red), leaving the disk $t_{\rm leave}$ (orange), and stellar wind collisions with CO $t_{\rm hit}$ (blue) Vs distance to the gas. The fiducial model labelled ``young system'' has $M_\star=1 M_\odot$, $v_{\rm SW}=100$ km/s, and $\Omega_\star=10$ $\Omega_\odot$. The timescale $t_{\rm ion}$ for SW protons is shown for a young system but it is always an order of magnitude larger than $t_{\rm hit}$. }
\end{figure}

Figure~\ref{fig2} shows $t_{\rm ion}$, $t_{\rm leave}$ and $t_{\rm hit}$ as a function of $r$, the radial distance to the central star. We notice that the SW collisional timescale $t_{\rm hit}$ becomes smaller with decreasing age or increasing mass loss rate, which are expected to be at their highest when the systems are young. We also find that increasing collision velocity will increase $t_{\rm hit}$ by a small amount. 
We see that at the typical location of debris disks (between $\sim$30-150 au), the smallest timescales are always $t_{\rm leave}$ and $t_{\rm hit}$, meaning that before CO, or O have time to ionise, they are pushed outside of the main disk (unless they are already ionised when released). However we note that ionisation via SW protons of CO is the most efficient and some CO$^+$ may be present. For completion, we note that the photodissociation of CO takes roughly 120 yr at large distances from the star \citep[owing to the ISM photons,][]{2009A&A...503..323V} and for other important timescales, refer to Tables C1 and C2 in \citealt{2021A&A...653L..11K}. Most of the gas close to the belt is then expected to be neutral and molecular, which could ease detections with ALMA targeting CO lines.

We do not rule out that for a specific very young system that is still emitting large amounts of FUV photons, the ionisation timescale may become smaller than calculated here from the standard \citet{2003IAUS..210P.A20C}'s stellar spectra and in this case ionised species could be present as well as individual C or O atoms (which may also get ionised) rather than CO close to the belt, and CO$^+$ may start to dominate. It could also be that in old systems the stellar mass loss rate becomes too small and gas has time to ionise and/or photodissociate before it gets pushed away by SW protons. However, since most observations are for young systems currently, this case is not often expected but might still drive our observing strategy to detect belt winds. We still recommend to perform these timescale comparisons when studying a specific system for assessing the state of the gas given its specific stellar spectra, angular velocity, mass loss rate.

\section{Retrieving stellar wind characteristics from gaseous belt outflows}\label{velanddens}

Here, we describe how we would retrieve the SW density and velocity around main sequence stars thanks to, e.g., ALMA observations of CO or neutral carbon gas. It is important to find new techniques to do that because currently we only have a dozen of measurements for main-sequence stars \citep{2011ApJ...741...54C} but none for A stars \citep{1992A&A...257..663L,2014A&A...564A..70K}. 

The first thing to do is to retrieve the carbon or CO density in the belt (with a Keplerian motion) from observations (or put an upper limit to it if the wind dominates and hide the Keplerian component), which should be easy because in these low gas mass systems, lines are optically thin and we expect the gas to be close to the purely (non-LTE) radiative regime where the line intensity does not depend on temperature. Hence, we can retrieve the gas mass and density $n_{\rm dobs}$. If the wind (moving spherically outwards) is also detected, we can retrieve its density $n_{\rm wobs}$.  In this case, we can also derive the mean gas velocity $\langle v_{\rm wobs} \rangle$ from the associated redshift of the wind compared to the star velocity and after correcting for potential orientation effects.

Assuming steady-state and using Eq.~\ref{ngasss3} with $n_{\rm dobs}$, we can go back to the factor ${\rm \dot{M}_g}/{\rm \dot{M}}_{\rm SW}$. If the windy part of the gas is not detected then using a model for the gas production rate such as described by Eq.~\ref{masslossrateeq} \citep{2017MNRAS.469..521K}, we can retrieve the stellar mass loss rate ${\rm \dot{M}}_{\rm SW}$. If the wind is detected then we can use the simple one zone model presented in the paper to get some first estimates of the stellar wind velocity assuming that $v_{\rm SW} \approx \mu \langle v_{\rm wobs} \rangle$. We can also retrieve a value of ${\rm \dot{M}}_{\rm SW}$, which does not depend on ${\rm \dot{M}_g}$ (hence independent of a gas release model) using Eq.~\ref{nwtogratio}, i.e. by computing the ratio between $n_{\rm wobs}$ and $n_{\rm dobs}$.


The first unambiguous detection of a belt wind would have the beneficial side effect of allowing us to improve our model further. Such an improved model (that is beyond the scope of this paper) would extend our one zone approach to a numerical multi-zone model using monte carlo simulations to compute the outcome of the different collisions with SW protons and reach a steady-state. This could be particularly useful for resolved observations. For short collisional timescales with the SW, there could be multiple collisions before reaching the end of the SW bubble (called heliosphere in our Solar System and located at $\sim$ 150 au, \citealt{2020NatAs...4..675O}). One could also include collisions with protons from the local interstellar medium, whose density is around 0.1 cm$^{-3}$ \citep{1997A&A...317..193I}, when the gas reaches beyond the SW bubble. The end velocity at large distances would then be close to that of the protons from the ISM. Finally, the gas state (owing to ionisation, photoionisation, ...) could also be followed more accurately coupling the code to a PDR-like model to make some predictions for different lines.



\section{Prediction of the wind profiles at large distances and observations with ALMA}\label{ALMA}

\subsection{The wind profile at large distances from the belt}

Our approach assumes that, after a collision, gas particles exit the disk and they do not get hit by another proton afterwards. First, we create a 3-D cartesian grid and we will call production cells the space between 60 and 90 au, and with $z$ lower than the scale height, i.e. where the planetesimal belt is located. Exterior cells (or wind cells) then refer to all the other cells. Our approach is simple : each production cell (x,y,z) in the disk produces a flux of scattered molecules in a cone, directed in the radial direction. The emitted flux density and velocity of the scattered molecules depends on the angle of collision $\psi$ we defined in Appendix \ref{coloutcome}. For each exterior cell, we add the contributions, to both the wind density and its velocity distribution from all production cells whose post-collision gas outflow cone crosses this exterior cell. 

In practice, we assume that the gas velocity is constant after impact with a proton and that the density decreases as $r_d^{-2}$ beyond the production cell, $r_d$ being the distance between the production and the exterior cells considered. We run through each production cell and consider all exterior cells that it can target (i.e. all cells beyond the half sphere or cone of influence) and add the contribution weighted by $r_d^{-2}$ to the density it had when emerging the belt. We also calculate the angle between the production and exterior cells and multiply the contribution in density by $1/\cos{\psi}$ because gas particles leaving their cells at high angles (e.g. $\psi$ is close to $\pi/2$ for edge-on collisions with protons) will move more slowly than for head-on collisions. However, there will also be less collisions at high angles because high impact angles cover less impact parameters than low ones and we add a $\cos{\psi}$ contribution to account for that effect, thus cancelling the previous effect. We proceed in the same way to get the contribution in velocity from all interior cells and calculate the velocity direction by computing the normalised vector between the two cells. We also weight the different contributions by the density and account for the $\cos{\psi}$ factor because head-on collisions will lead to a faster moving wind. We then average the different velocity vector contributions coming from all production cells and consider the final velocity vector as being representative of what a mean radial velocity offset will look like in real observations. The density and velocity fields are plotted in Fig.~\ref{figwind} and used to make the synthetic observations presented in the paper.

\subsection{ALMA synthetic images for future observations}


To make detailed predictions for detectability of a wind with future observations, we focus on the TWA 7 system and follow four steps. 

First, we define the density and velocity structure of the CO disk and the wind component. For the CO disk component, we assume that it has a constant number density radially between 60 and 90 au, and a (radially constant) scale height of 6 au. The velocity field of this component is assumed to be in Keplerian rotation around a star of 0.51 M$_{\odot}$. For the wind component, we assume that its density $n_{\rm w}$ simply follows $n_{\rm w}/n_{\rm d}=0.1$ at the disk location (worst case scenario). At radii and vertical heights larger than the disk's, we assume the density and velocity structure described at the beginning of this section.

Second, we calculate the excitation of the CO gas assuming it is out of LTE, and purely in the radiation-dominated (low density) regime. This is justified by the low densities expected in an environment largely devoid of species other than CO. At each position within our model, we calculate CO level populations using non-LTE codes including the effect of UV/IR fluorescence due to absorption of UV/IR stellar/interstellar radiation \citep{2015MNRAS.447.3936M,2018ApJ...853..147M}. The stellar and interstellar radiation components are as described in \citet{2019AJ....157..117M}, with the stellar contribution rescaled with the inverse of the gas stellocentric distance. To keep our feasibility test as realistic as possible using measured line intensities, we target the same CO transition (J=3-2) detected by existing ALMA observations \citep{2019AJ....157..117M}.

Third, we use our 3D density, velocity, and excitation structure in our model to carry out radiative transfer using the RADMC-3D code \citep{2012ascl.soft02015D}. For the temperature structure, for simplicity we assume radius-dependent blackbody temperatures appropriate for the luminosity of the host star; we note that these temperatures do not affect excitation in the radiation-dominated regime, but only the intrinsic CO J=3-2 line widths. This produces a spectro-spatially resolved cube of CO J=3-2 line intensities. The output cube has a pixel size of 0.1$"$, extending spatially out to $\pm$6.4$"$ from the star. The cube has a channel size of 244.14 kHz (0.21 km/s at the line location of 345.796 GHz), and extends out to $\pm21$ km/s from the stellar velocity. Note that we rescale the input CO mass to produce a peak CO line intensity (after spatially integrating over the entire model) of 60 mJy in a 244.14 kHz channel, to ensure consistency with the existing ALMA dataset \citep[see e.g.][Fig. 1, top spectrum]{2019AJ....157..117M}. This leads to a model with a characteristic CO number density of $\sim$100 cm$^{-3}$ within the disk component. We note that this is an upper limit of the density in the disk because we assumed a purely radiation-dominated regime but a few colliders could bring it closer to LTE in a regime where the gas density would not need to be as high to explain the CO detection around TWA 7. For instance, in LTE we find that a gas density of 0.08 cm$^{-3}$ in the disk is enough to explain the CO J=3-2 detection. The real density therefore must lie in between but our synthetic images do not change much between the two regimes because of the rescaling to get the right intensity and our predictions remain valid whatever the amount of colliders.

\begin{figure}
   \centering
   \includegraphics[width=9.cm]{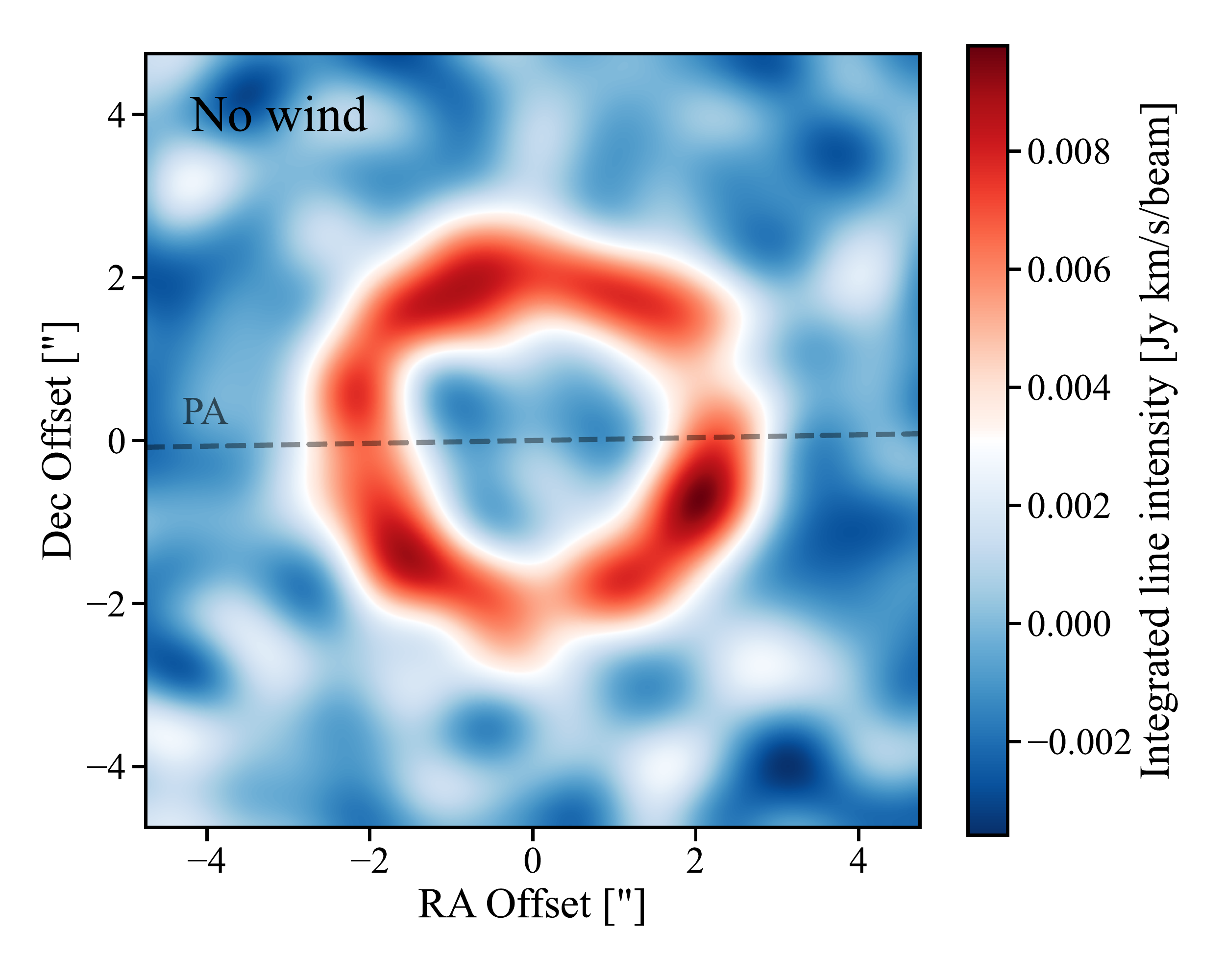}
      \includegraphics[width=9.cm]{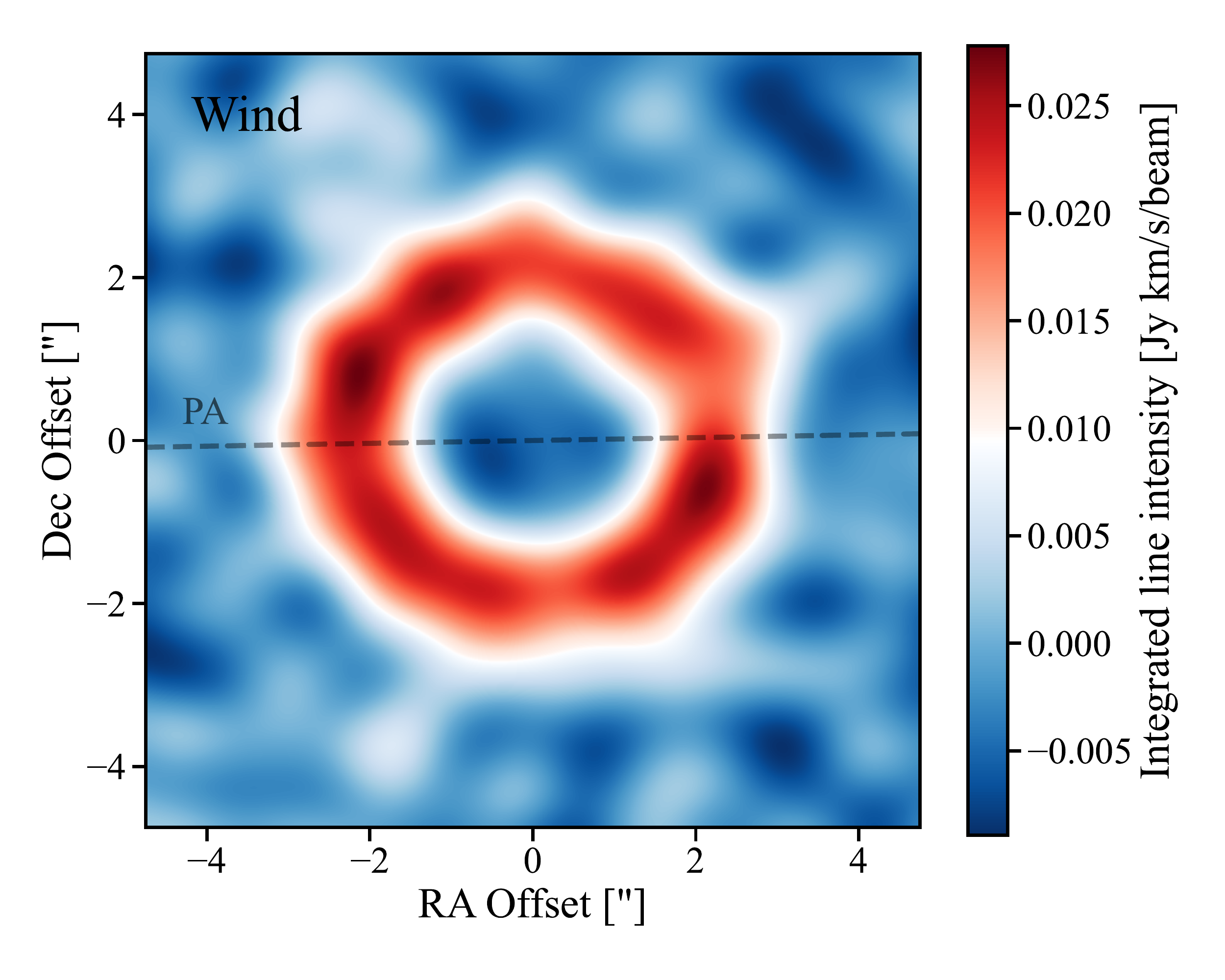}

   \caption{\label{figm0} Moment-0 images for the no wind (top) and wind (bottom) cases. They look very similar given the low spatial resolution and sensitivity level and we need to look at the velocity component of the gas to clearly see the difference between the wind and Keplerian models. }
\end{figure}

Finally, we produce synthetic ALMA visibilities using the \textit{simobserve} task within the CASA software v6.4 \citep[][]{2007ASPC..376..127M}. Due to the low expected CO surface brightness of CO emission, we simulate 345 GHz (Band 7) observations in the most compact array configuration (C-1, baselines from 15 m to 161 m), leading to a beam size of $1.06"\times1.25"$, corresponding to $\sim$40 au at the distance of the star. We simulate enough repetitions to reach a cube sensitivity of 1.5 mJy/beam in every 244.14 kHz channel, a factor of 4.6 deeper than existing observations. Visibilities are imaged to produce line cubes using the \textit{tclean} CASA task, using natural weighting and standard Hogbom deconvolution.
To produce moment-0 images (Fig.~\ref{figm0}), we spectrally integrate the cube over velocities where the CO J=3-2 line is significantly detected ($\pm6.7$ km/s), whereas to produce 1D spectra we spatially integrate in a radial region between 1.2$"$ (40 au) and 3.2$"$ (110 au) from the central star. Finally, to produce moment-1 maps (Fig.~\ref{figm1}), we select the same spatial region, and calculate the velocity centroid of emission in the 1D spectrum within each of the spatially selected pixels. The creation of moment maps is implemented using the \textit{bettermoments} package \citep[][]{2018RNAAS...2..173T}. We also produce the spectra and PV diagrams as shown in Fig.~\ref{figs} and \ref{figpv}, respectively.

We find that the wind can be easily distinguished from a Keplerian gas disk from the spectrum and moment-0 image because we can see that the spectrum does not fit that expected from a Keplerian profile given the moment-0 image. However, the easiest way to clearly see that gas is wind-like is to look at the moment-1 image, which shows a clear difference to a Keplerian pattern in that the gas velocity is perpendicular to the disk position angle (and not aligned with it as expected for a Keplerian gas disk). This difference in the velocity profile is also clearly visible in the respective PV diagrams and would allow observers to spot a windy gas component straight away.

\section{Gas sample used}\label{obs}

In Table~\ref{tab:table1}, we list the references from which we obtained the value of $n_{\rm d}$ and $\Delta R$ we use to plot Figure \ref{fig1}. The sample includes both high-gas-density systems and low-density ones. It is not complete by any means but serves to guide our eyes on the different plots. The first column gives the star's name of the system with gas detected. The second column provides the method to calculate the gas density (gas mass, gas surface density) or ``reference'' if it is already provided as such in the reference. The third column is for the gas density that is either calculated from the mass, when provided, or directly taken from the papers listed in the Ref. column. The fourth column is for the belt's width $\Delta R$ and the last column provides the references we used to get both the gas amount and its width. Note that if there are two references, the first one if for the estimation of the amount of gas and the other for $\Delta R$. We take the literature values as is and do not try to remodel everything homogeneously, which means that horizontal (width) as well as vertical (density) error bars may be quite large because sometimes the CO is not well resolved and/or the excitation/optical depth conditions are not very well known. Our plot rather shows the trend and that some disks may be in the WIND regime but dedicated studies should be carried out for each system to find out for sure.


\newcommand\kq{{\citet{2021A&A...653L..11K}}}

\begin{table*}
  \begin{center}
    \caption{Parameters used in Fig.~\ref{fig1} with their references. Reference in the method column means that we obtained the gas density straight from the indicated reference listed in the last column.}
    \label{tab:table1}
    \begin{tabular}{|c|c|c|c|c|} 
      \hline

      \textbf{Star's name} & \textbf{Method} & \textbf{Gas density (cm$^{-3}$)} & \textbf{$\Delta R$ (au)} & \textbf{Ref.} \\
            \hline

      $\beta$ Pic & reference  & 350 & 100 & \citet{2016MNRAS.461..845K,2014Sci...343.1490D} \\
      \hline
     Fomalhaut &  reference & 0.02-0.75 & 13.6 & \citet{2017ApJ...842....9M,2017ApJ...842....8M} \\
      \hline
        NO Lup & Gas mass & 9 & 55 & \citet{2021MNRAS.502L..66L} \\
      \hline
       TWA 7 & Gas mass & 0.2-20 & 80 & \citet{2019AJ....157..117M} \\
      \hline
        HD 129590 & Gas surface density & 30 & 60 & \citet{2020MNRAS.497.2811K} \\
      \hline
        HD 21997 &  reference & 800-10000 & 88 & \citet{2020ApJ...905..122H,2013ApJ...777L..25M} \\
      \hline
        49 Ceti & reference & 100-800 & 215 & \citet{2020ApJ...905..122H,2017ApJ...839...86H} \\
      \hline
      The Sun (Kuiper belt) &  reference & $3 \times 10^{-7}$ & 10 & \kq  \\
      \hline

     





         \end{tabular}
  \end{center}
\end{table*}

\end{appendix}

\label{lastpage}

\end{document}